\documentclass[smallextended]{svjour3} 

\usepackage{cite}
\usepackage{amsmath,amssymb,amsfonts}
\usepackage{algorithmic}

\usepackage{graphicx}
\usepackage{epstopdf}

\usepackage{textcomp}
\usepackage{xcolor}
\usepackage{mathtools}
\usepackage{bbm}
\usepackage{comment}
\usepackage{dsfont}
\usepackage{caption}
\usepackage{tabularx,booktabs}
\usepackage{slashbox}
\usepackage{booktabs,caption}
\usepackage[flushleft]{threeparttable}
\usepackage{multirow}

\usepackage{amsmath}
\usepackage{color}
\usepackage{comment}
\usepackage{amsmath}
\usepackage{appendix}
\usepackage{svg}

\usepackage{adjustbox}

\DeclareMathOperator*{\argmin}{arg\,min}

\usepackage{overpic}

\newcommand{\piclabC}[2]{
\begin{overpic}[width=0.95\linewidth]{#1}
 \put (3,98) {\large \textsf{#2}}
\end{overpic}
}

\usepackage{tabularx}
\usepackage{subcaption}
\newcolumntype{C}[1]{>{\centering\arraybackslash}p{#1}}

\newcommand\norm[1]{\left\lVert#1\right\rVert}

\begin{document}

\title{Temporal mixture ensemble for probabilistic forecasting of intraday volume in cryptocurrency exchange markets}

\author{Nino Antulov-Fantulin* \and
        Tian Guo* \and
        Fabrizio Lillo}

\titlerunning{Temporal mixture ensemble for cryptocurrency intraday volume forecasting}

\institute{*Shared first authorship.\\
N. Antulov-Fantulin \at
              ETH Zurich \at
              Aisot GmbH, Zurich, Switzerland
             \and 
              T. Guo \at
              RAM Active Investments, Switzerland \\
              Work done when at ETH Zurich 
              \and
              F. Lillo \\
              Department of Mathematics,University of Bologna\\ Scuola Normale Superiore, Pisa, Italy\\
              \email{fabrizio.lillo@unibo.it} 
}

\date{Received: date / Accepted: date}
    

\maketitle

\begin{abstract}
We study the problem of the intraday short-term volume forecasting in cryptocurrency exchange markets. 
The predictions are built by using transaction and order book data from different markets where the exchange takes place.
Methodologically, we propose a temporal mixture ensemble, capable of adaptively exploiting, for the forecasting, different sources of data and providing a volume point estimate, as well as its uncertainty. 
We provide evidence of the outperformance of our model by comparing its outcomes with those obtained with different time series and machine learning methods. Finally, we discuss the predictions conditional to volume and we find that also in this case machine learning methods outperform econometric models.  
\end{abstract}

\section{Introduction} 

Cryptocurrencies recently attracted massive attention from public and researcher community in several disciplines such as finance and economics~\cite{urquhart2016inefficiency,Bolt2016,cheah2015speculative,Chu2015,BouchaudBTC,ciaian2016economics}, computer science~\cite{Ron2013BTC,jang2018empirical,amjad2017trading,alessandretti2018machine,guo2018bitcoin} or complex systems~\cite{Garcia2015,wheatley2019bitcoin,gerlach2019dissection,antulov2018inferring, KondorBTC,ElBahrawy2017}. 
It originated from a decentralized peer-to-peer payment network~\cite{Nakamoto2008}, relying on cryptographic methods~\cite{bos2014elliptic,mayer2016ecdsa} like elliptical curve cryptography and the SHA-256 hash function.
When new transactions are announced on this network, they have to be verified by network nodes and recorded in a public distributed ledger called the blockchain \cite{Nakamoto2008}.
Cryptocurrencies are created as a reward in the verification competition (see Proof of work~\cite{jakobsson1999proofs}), in which users offer their computing power to verify and record transactions into the blockchain.
Bitcoin is one of the most prominent decentralized digital cryptocurrencies and it is the focus of this paper, although the model developed below can be adapted to other cryptocurrencies with ease, as well as to other "ordinary" assets (equities, futures, FX rates, etc.). 

The exchange of Bitcoins with other fiat or cryptocurrencies takes place on exchange markets, which share some similarities with the foreign exchange markets \cite{BAUMOHL2019363}.
These markets typically work through a continuous double auction, which is implemented with a limit order book mechanism, where no designated dealer or market maker is present and limit and market orders to buy and sell arrive continuously. Moreover, as observed for traditional assets, the market is {\it fragmented}, i.e. there are several exchanges where the trading of the same asset, in our case the exchange of a cryptocurrency with a fiat currency,  can simultaneously take place.
\begin{figure}[!htbp]
     \centering
     \includegraphics[width=0.9\textwidth]{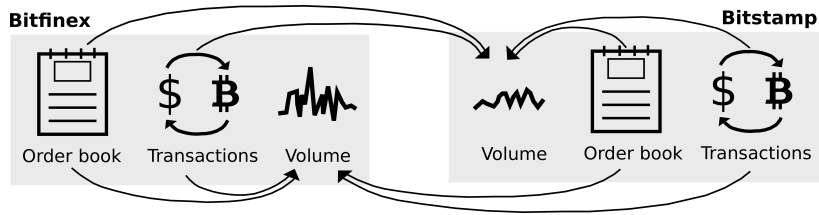}
     \caption{Illustration of the probabilistic volume predicting in the multi-source data setting of Bitfinex and
     Bitstamp markets.
     The left and right panel respectively depict the order book and transaction data of each market.
     The arrows represent the data used to model the volume of each market.
     Note that the volume information is implicitly contained in the transaction data of one market and thus there is no arrow linking the volumes from the two markets.}
     \label{fig:multi_src}
\end{figure}

The automation of the (cryptocurrency) exchanges lead to the increase of the use of automated trading \cite{algo1,algo2} via different trading algorithms. An important input for these algos is the prediction of future trading volume. This is important for several reasons. First, trading volume is a proxy for liquidity which in turn is important to quantify transaction costs. Trading algorithms aim at minimizing these costs by splitting orders in order to find a better execution price \cite{frei2015optimal, barzykin2019optimal} and the crucial part is the decision of when to execute the orders in such a way to minimize market impact or to achieve certain trading benchmarks (e.g. VWAP)
\cite{brownlees2010intra, satish2014predicting, chen2016forecasting,bialkowski2008improving, calvori2013go,kawakatsu2018direct}. Second, when different market venues are available, the algorithm must decide where to post the order and the choice is likely the market where more volume is predicted to be available.
Third, volume is also used to model the time-varying price volatility process, whose relation is also known as “Mixture of Distribution Hypothesis" \cite{andersen1996-MDH}.

In this paper, we study the problem of intraday short-term volume prediction on multi-market of cryptocurrency, as is shown in Fig.~\ref{fig:multi_src}, intending to obtain not only point estimate but also the uncertainty on the point prediction~\cite{Chu2015, urquhart2016inefficiency, Katsiampa2017, balcilar2017can}.
Moreover, conventional volume predictions focuses on using data or features from the same market. 
Since cryptocurrency markets are traded on several markets simultaneously, it is reasonable to use cross-market data not only to enhance the predictive power, but also to help understanding the interaction between markets.
In particular, we investigate the exchange rate of Bitcoin (BTC) with a fiat currency  (USD) on two liquid markets: Bitfinex and Bitstamp.
The first market is more liquid than the second, since its traded volume in the investigated period from June 2018 to November 2018 is $2.5$ times larger\footnote{Recently, there have been few reports that are showing fake reported volume for certain Bitcoin exchange markets. In this paper, we investigate Bitcoin exchange markets that have regulatory status~\cite{hougan2019SEC} either with the Money Services Business (MSB) license or BitLicense from the New York State Department of Financial Services, and have been independently verified to report true values.}. 
Thus one expects an asymmetric role of the past volume (or other market variables) of one market on the prediction of volume in the other market.

Specifically, the contribution of this paper can be summarized as follows:
\begin{itemize}
    \item We formulate the cross-market volume prediction as a supervised multi-source learning problem.
    We use multi-source data, i.e. transactions and limit order books from different markets, to predict the volume of the target market.
    \item We propose the \textbf{T}emporal \textbf{M}ixture \textbf{E}nsemble (TME), which models individual source's relation to the target and adaptively adjusts the contribution of the individual source to the target prediction.
    \item By equipping with modern ensemble techniques, the proposed model can further quantify the predictive uncertainty consisting of the epistemic and aleatoric components, on the predicted volume.
    \item As main benchmarks for volume dynamics, we use different time-series and machine learning models (clearly with the same regressors/features used in our model). 
    We observe that our dynamic mixture ensemble is often having superior out-of-sample performance on conventional prediction error metrics e.g. root mean square error (RMSE) and mean absolute error (MAE). 
    More importantly, it presents much better calibrated results, evaluated by metrics taking into account predictive uncertainty, i.e. normalized negative log-likelihood (NNLL), uncertainty interval width (IW).
    \item We discuss the prediction performance conditional to volume. Since our choice of modeling log-volume is tantamount to considering a multiplicative noise model for volumes, when using relative RMSE and MAE machine learning methods outperforms econometric models in providing more accurate forecasts. 
\end{itemize}

The paper is organized as follows: in Sec.~\ref{sec:data} we present the investigated markets, the data, and the variables used in the modeling. In Sec.~\ref{sec:model} we present our benchmark models. 
In Sec.~\ref{sec:exp} we present our empirical investigations on the cryptocurrency markets for the prediction of intraday market volume. Finally, Sec.~\ref{sec:conclusion} presents some conclusions and outlook for future work. Most of the technical description of models and algorithms, as well as some additional empirical results, are presented in an appendix. 
\section{Multiple market cryptocurrency data}\label{sec:data}

Our empirical analyses are performed on a sample of data over the period from May 31, 2018 9:55pm (UTC) until Spetember 30 2018 9:59pm (UTC) from two exchange markets, Bitfinex\footnote{https://www.bitfinex.com} and Bitstamp\footnote{https://www.bitstamp.net}, where Bitcoins can be exchanged with US dollars.
These markets work through a limit order book, as many conventional exchanges.
For each of the two markets we consider two types of data: transaction data and limit order book data.


From \textbf{transaction data} we extract the following features on each 1-min interval:

\begin{itemize}
    \item \textit{Buy volume} - number of BTCs traded in buyer initiated transactions 
    \item \textit{Sell volume} - number of BTCs traded in seller initiated transactions
    \item \textit{Volume imbalance} - absolute difference between buy and sell volume
    \item \textit{Buy transactions} - number of executed transactions on buy side
     \item \textit{Sell transactions} - number of executed transactions on sell side
     \item \textit{Transaction imbalance} - absolute difference between buy and sell number of transactions
\end{itemize}
We remind that a buyer (seller) initiated transaction in a limit order book market is a trade where the initiator is a buy (sell) market order or a buy (sell) limit order crossing the spread.

From \textbf{limit order book data} we extract the following features each minute~\cite{OB_rev_Porter2013, Lillo2016}:

\begin{itemize}
    \item \textit{Spread} is the difference between the highest price that a buyer is willing to pay for a BTC (bid) and the lowest price that a seller is willing to accept (ask).  
    \item \textit{Ask volume} is the number of BTCs on the ask side of order book.
     \item \textit{Bid volume} is the number of BTCs on the bid side of order book.
    \item \textit{Imbalance} is the absolute difference between ask and bid volume. 
    \item \textit{Ask/bid Slope} is estimated as the volume until $\delta$ price offset from the best ask/bid price.
     $\delta$ is estimated by the bid  price at the order that has at least 1$\%$, 5$\%$ and 10 $\%$ of orders with the highest bid price. 
    \item \textit{Slope imbalance} is the absolute difference between ask and bid slope at different values of price associated to $\delta$. $\delta$ is estimated by the bid  price at the order that has at least 1$\%$, 5$\%$ and 10 $\%$ of orders with the highest bid price.
    \end{itemize}
\begin{figure}[!htbp]
     \centering
     \begin{subfigure}[b]{0.9\textwidth}
         \centering
         \includegraphics[width=1.0\textwidth]{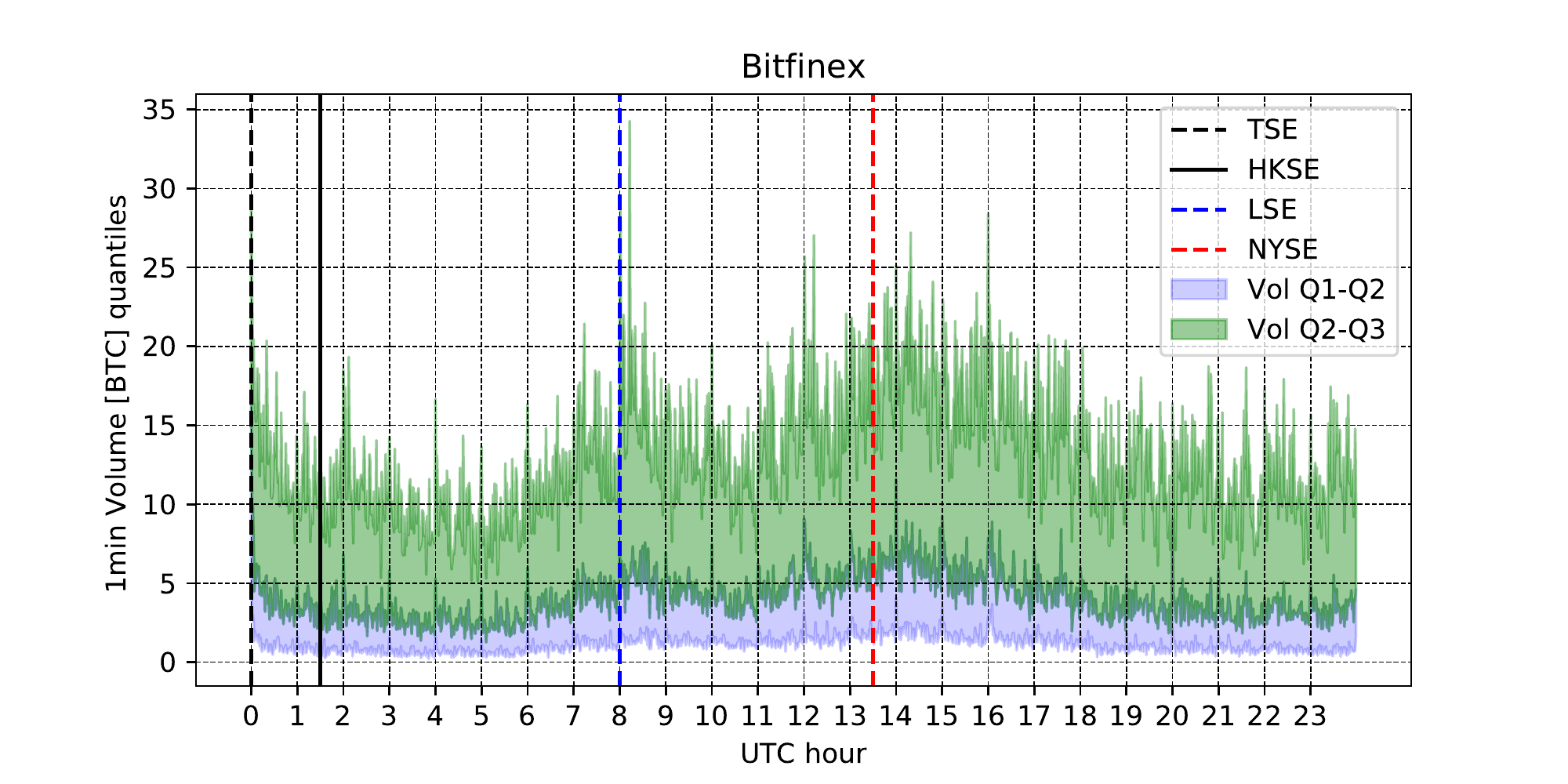}
     \end{subfigure}
     \hfill
     \begin{subfigure}[b]{0.9\textwidth}
         \centering
         \includegraphics[width=1.0\textwidth]{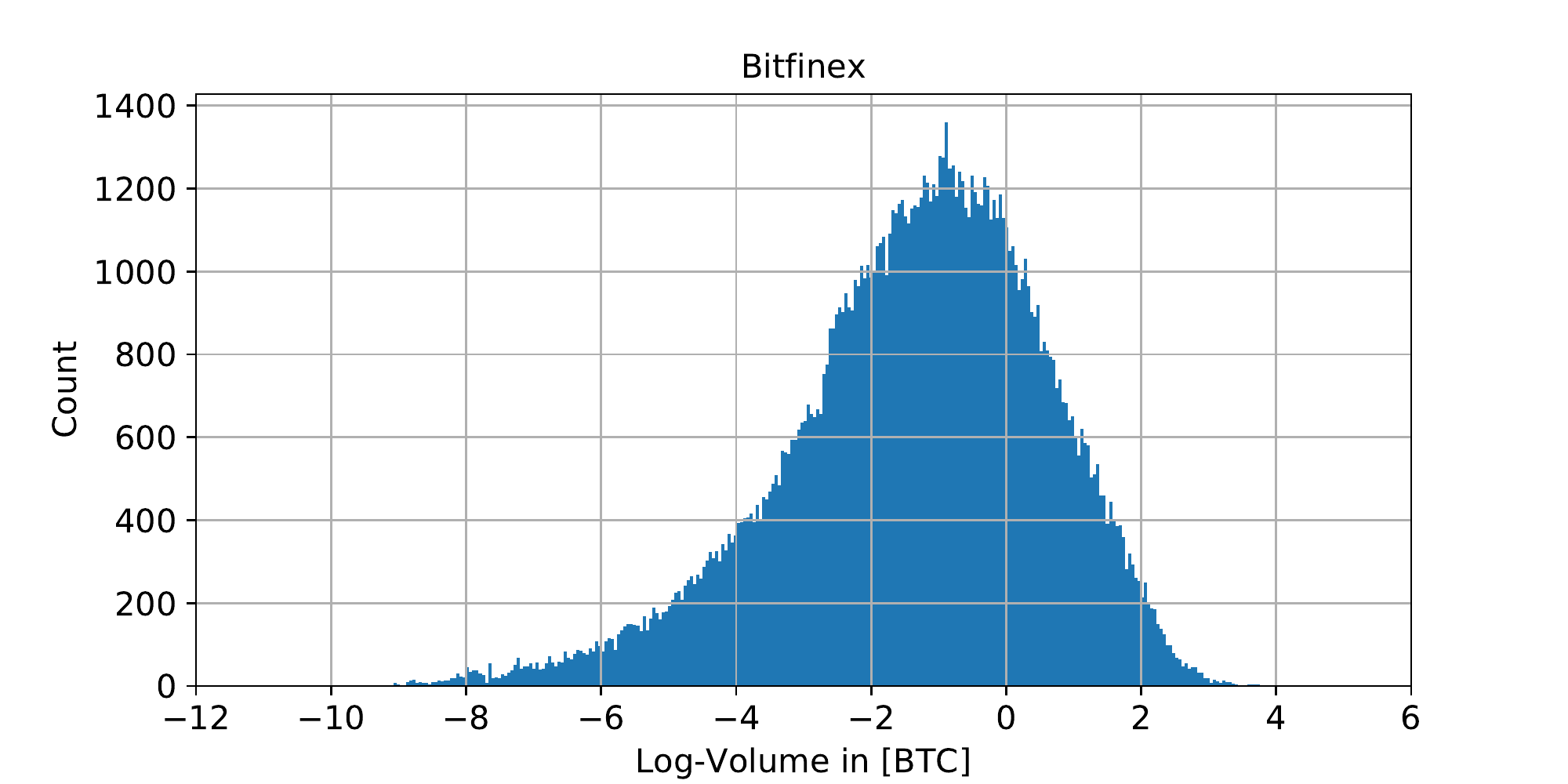}
     \end{subfigure}
     \hfill
     \begin{subfigure}[b]{0.9\textwidth}
         \centering
         \includegraphics[width=1.0\textwidth]{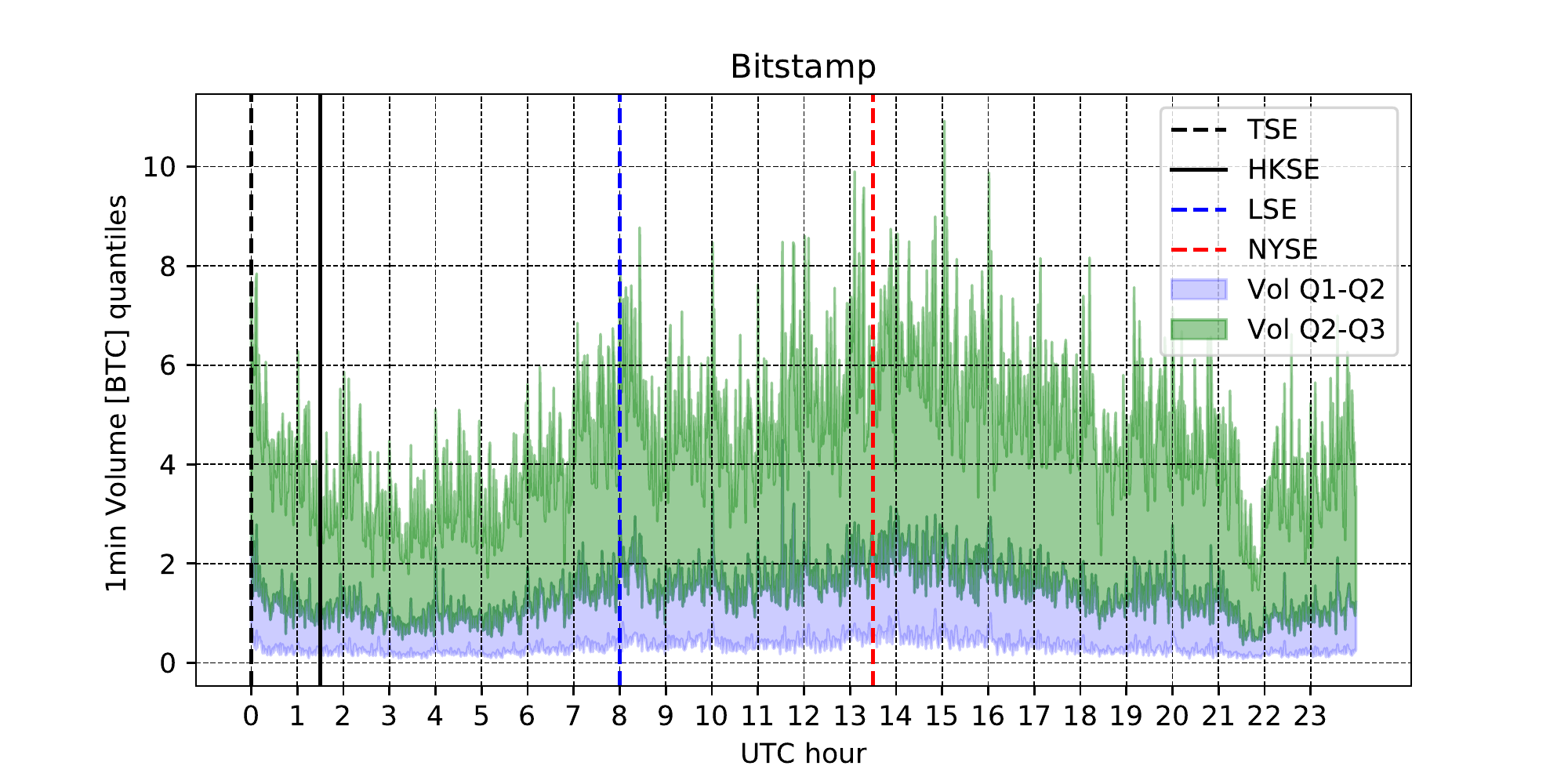}
     \end{subfigure}
     \hfill
     \begin{subfigure}[b]{0.9\textwidth}
         \centering
         \includegraphics[width=1.0\textwidth]{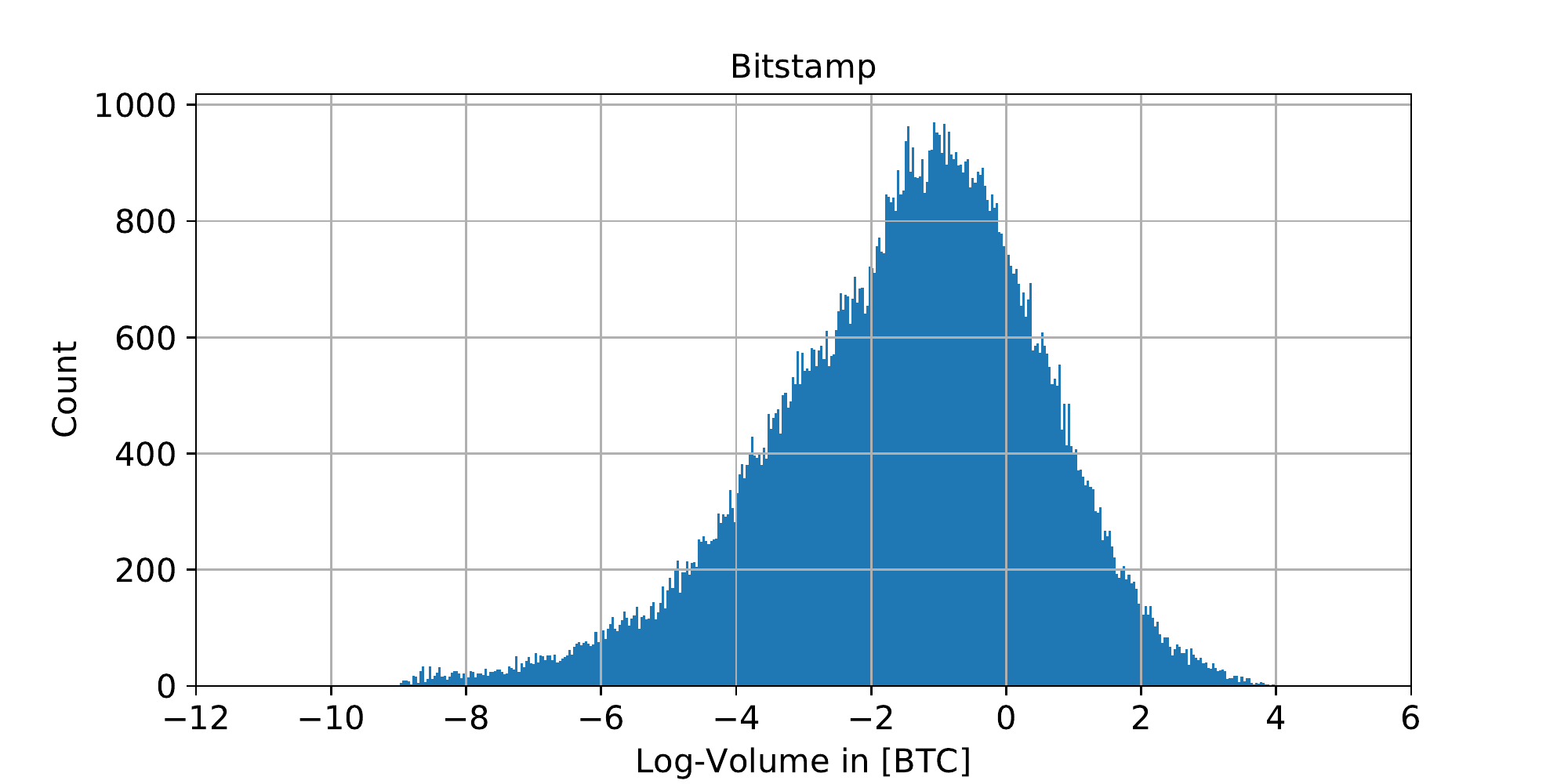}
     \end{subfigure}
     \hfill
\caption{\textbf{1st and 3rd panel}: interquartile range $Q_1-Q_2$ and $Q_2-Q3$ of intraday 1-min transaction volume of BTC/USD rate on Bitfinex (1st panel) and Bitstamp (3rd panel) market, along with openings of major exchanges (TSE, HKSE, LSE, NYSE) that are denoted with vertical lines.
\textbf{2nd and 4th panel}: histogram of deseasonalized log volume at 1-min  resolution. The statistics of these variable are:
Bitfinex: mean = -1.3627, variance = 3.7658, skewness = -0.6336, and  kurtosis = 0.6886;
Bitstamp: mean = -2.4224, variance = 3.9894, skewness = -0.5307, kurtosis = 0.4114.
 }
\label{fig:volume_stat}
\end{figure}

The target variable that we aim at forecasting is the trading volume of a given target market including both buy and sell volume.
In the proposed modeling approaches (described in Sec.~\ref{sec:model}) we consider different data sources at each time can affect the probability distribution of trading volume in the next time interval in a given market.
As is illustrated in Fig.~\ref{fig:multi_src}, given the setting presented above, there are four sources, namely one for transaction data and one for limit order book data for the two markets.

Before going into the details of models, in order to choose the appropriated variable to investigate, we visualize some characteristics of the data in Fig.~\ref{fig:volume_stat}.
In the 1st and 3rd panel, we show the  quantiles of intraday 1-min trading volume of BTC/USD rates, for the two different markets. 
We also show as vertical lines the opening times of four major stock exchanges. Although we do not observe abrupt changes in volume distribution (possibly because cryptocurrency exchanges are weakly related with stock exchanges), some small but significant intraday pattern \cite{andersen1997intraday} is observed. 

For this reason, we pre-process the raw volume data to remove intraday patterns as follows.
Let us denote $v_{t}$ the volume traded at the time $t$, in units of Bitcoins.
${I(t)}$ is a mapping function of the time $t$, which returns the intraday time interval index of $t$.
We use $a_{I(t)}$ to represent the average of volumes at the same intraday index $I(t)$ across days.
Next, to remove intraday patterns, we process the raw volume by taking as modeled variable $y_t \triangleq \frac{v_{t}}{a_{I(t)}}$.
In practise, in order to avoid leaking information to the training phase, $a_{I(t)}$ is calculated only based on the training data, and shared to both validating and testing data.

The  histogram of $\log y_t$ for the two markets\footnote{The frequency of time intervals with zero volume and how we handle them is detailed in Section \ref{sec:data}.} is shown in the 2nd and 4th panel of Fig.~\ref{fig:volume_stat} along with first four cumulants of empirical distribution of log-volumes. We observe that the distribution is approximately normal, even if a small negative skew is present. For this reason, as done in the literature (see for example \cite{bauwens2008moments}, our modeling choice is to consider $y_t$ as log-normally distributed.

\section{Models}\label{sec:model}

Econometric modeling of intra-daily trading volume relies on a set of empirical regularities \cite{brownlees2010intra, satish2014predicting, chen2016forecasting} of volume dynamics. These include fat tails, strong persistence and an intra-daily clustering around the "U"-shaped periodic component.
Brownlees et al. \cite{brownlees2010intra} proposed Component Multiplicative Error Model (CMEM), which is the extension of Multiplicative Error Model (MEM) \cite{engle2002new}.
The CMEM volume model has a connection to the component-GARCH \cite{ComponentGARCH} and the periodic P-GARCH \cite{PGARCH}.
Satish et al. \cite{satish2014predicting}, proposed four-component volume forecast model composed of:
(i) rolling historical volume average,
(ii) daily ARMA for serial correlation across daily volumes, 
(iii) deseasonalized intra-day ARMA volume model and
(iv) a dynamic weighted combination of previous models.
Chen et al. \cite{chen2016forecasting}, simplify the multiplicative volume model \cite{brownlees2010intra} into an additive one by modeling the logarithm of intraday volume with the Kalman filter.

\subsection{Problem setting}

In this paper, we focus on the probabilistic forecasting including both mean and variance of the intraday volume in the multi-source data setting. We propose the use of TME, presented below, and we benchmark its performance against two econometric baseline models (ARMA-GARCH and ARMAX-GARCH) and one Machine Learning baseline model (Gradient Boosting Machine). 

As mentioned above, we assume $y_t$ follows the log-normal distribution~\cite{bauwens2008moments} and thus all the models used in the experiments will be developed to learn the log-normal distribution of $y_t$.
However, our proposed TME is flexible to be equipped with different distributions for the target, and the study in this direction will be the future work. 

When evaluating the performance of the forecasting procedure with real data experiments, we choose to use  evaluation metrics defined on the original volume $v_t$, since in real world application the interest is in forecasting volume rather than log-volume.
Finally, for understanding the performance of TME and Machine Learning and econometric baselines in different setups, we will evaluate them using three different time intervals of volumes, namely 1min, 5min and 10min.

Regarding the multi-source data setting, on one hand, it includes the features from the target market. 
This data is believed to be directly correlated with the target variable.
On the other hand, there is an alternative market, which could interact with the target market.
Together with the target market, the features from this alternative market constitute the multi-source data.

In this paper, we mainly focus on Bitfinex and Bitstamp markets.
For each market, we have the features from both transaction and order book data, thereby leading to $S = 4$ data sources.
In particular, we indicate with $\mathbf{x}_{s,t} \in \mathbb{R}^{d_s} $ the features from source $s$ at time step $t$, and ${d_s}$ the dimensionality of source data $s$.
Given the list of features presented in Sec.~\ref{sec:data}, we have $d_s=6$ when the source is transaction data in any market, while $d_s=13$ for order book data.
Then, these multi-source data will be used to model the volume of each market, as is shown in Fig.~\ref{fig:multi_src}.

\subsection{Overview of TME}

In this paper, we construct a Temporal Mixture Ensemble (TME), belonging to the class of of mixture models \cite{waterhouse1996bayesian, yuksel2012twenty, wei2007dynamic, bazzani2016recurrent, guo2019exploring}, which takes previous transactions and limit order book data \cite{OB_rev_Porter2013, Lillo2016} from multiple markets simultaneously into account. 
Though mixture models have been widely used in machine learning and deep learning 
\cite{guo2018bitcoin, schwab2019granger, kurle2019multi}, they have been hardly explored for prediction tasks in cryptocurrency markets.
Moreover, our proposed ensemble of temporal mixtures can provide predictive uncertainty of the target volume by the use of the Stochastic Gradient Descent (SGD) based ensemble technique \cite{lakshminarayanan2017simple, maddox2019simple, snoek2019can}.
Predictive uncertainty reflects the confidence of the model over the prediction. 
It is valuable extra information for model interpretability and reliability.
The model developed below is flexible to consume multi-source data of arbitrary number of sources and dimensionalities of individual source data.

In principle, TME exploits latent variables to capture the contributions of different sources of data to the future evolution of the target variable. 
The source contributing at a certain time depends on the history of all the sources. 

For simplicity, we will use one data sample of the target $y_t$ to present the proposed model.
In reality, the training, validation, and testing data contain the samples collected in a time period.
More quantitatively, the generative process of the target variable $y_t$ conditional on multi-source data $\{\mathbf{x}_{1,t}, \cdots, \mathbf{x}_{S,t}\}$ is formulated as the following probabilistic mixture process:
\begin{align}
\label{eq:joint}
\begin{split}
& p( y_t | \{\mathbf{x}_{1,<t}, \cdots, \mathbf{x}_{S,<t}\}, \Theta ) \\
= & \sum_{z_t = 1}^{S}
p_{\theta_s}(y_t | z_t = s, \mathbf{x}_{s, <t}) \cdot 
\mathbb{P}_{\omega}(z_t = s \, | \mathbf{x}_{1, <t}, \cdots, \mathbf{x}_{S, <t}). 
\end{split}
\end{align}
The latent variable $z_t$ is a discrete random variable defined on the set of values $\{1, \cdots, S\}$, each of which represents the corresponding data source. The quantity $p_{\theta_s}(y_t | z_t = s, \mathbf{x}_{s, <t})$ models the predictive probabilistic density of the target based on the historical data $\mathbf{x}_{s, <t}$ from a certain source $s$. 
The quantity\footnote{We indicate with $p_\theta$ probability densities and $\mathbb{P}_{\omega}$ probability mass functions.} $\mathbb{P}_{\omega}(z_t = s \, | \mathbf{x}_{1, <t}, \cdots, \mathbf{x}_{S, <t})$ characterize a time-varying categorical distribution dependent on multi-source data.
It adaptively adjusts the contribution of the data source specific density $p_{\theta_s}(y_t | z_t = s, \mathbf{x}_{s, <t})$ at each time step. Clearly, it holds
$\sum_{s = 1}^{S}\mathbb{P}_{\omega}(z_t = s \, | \mathbf{x}_{1, <t}, \cdots, \mathbf{x}_{S, <t}) = 1$. 
Finally, $\Theta \triangleq \{\omega, \theta_1, \cdots, \theta_S \}$ represents the parameters in data source specific components and the latent variable's probability function, and it will be learned in the training phase discussed below. 

\subsection{Model specification}

We now specify in detail the mathematical formulation of each component in the temporal mixtures.
Without loss of generality, we present the following model specification for the volume in cryptocurrency exchange of this paper's interest.

To specify the model, we need to define the predictive density function of individual sources, i.e. $p_{\theta_s}(y_t | z_t = s, \mathbf{x}_{s, <t})$ and the probability function of latent variable, i.e. $\mathbb{P}_{\omega}(z_t = s \, | \mathbf{x}_{1, <t}, \cdots, \mathbf{x}_{S, <t})$.
We make a general assumption for both these functions that data from different sources are taken within the same time window w.r.t. the target time step.
We denote by $h$ the window length, i.e. the number of past time steps which enter in the conditional probabilities. 
We assume that this value is the same for each source.
Eq.~\ref{eq:joint} is thus simplified as:
\begin{align}
\begin{split}
\sum_{z_t = 1}^{S}
& p_{\theta_s}(y_t | z_t = s, \mathbf{x}_{s, (-h,t)}) \cdot \mathbb{P}_{\omega}(z_t = s | \mathbf{x}_{1,  (-h,t)}, \cdots, \mathbf{x}_{S, (-h,t)}), 
\end{split}
\end{align}
where $\mathbf{x}_{s, (-h,t)}$ represents the data from source $s$ within the time window from $t-h$ to $t-1$ and $\mathbf{x}_{s, (-h,t)} \in \mathbb{R}^{d_s \times h}$. 

As for $p_{\theta_s}(y_t | z_t = s, \mathbf{x}_{s, (-h,t)})$, due to the non-negative nature of the target volume in cryptocurrency exchange and its statistical properties~\cite{bauwens2008moments}, we choose the log-normal distribution.
Given $\mathbf{x}_{s, (-h,t)} \in \mathbb{R}^{d_s \times h}$, we choose bi-linear regression to parameterize the mean and variance of the log transformed volume as follows:
\begin{gather}
\ln(y_t) \, | z_t \text{=} s, \mathbf{x}_{s,(-h, t)} \, \sim \, \mathcal{N}\big( \mu_{t,s}, \sigma_{t,s}^2) \\
\mu_{t,s} \triangleq L_{\mu, s}^{\top} \cdot \mathbf{x}_{s,(-h, t)} \cdot R_{\mu, s} + b_{\mu, s} \\
\sigma_{t,s}^2 \triangleq \exp(L_{\sigma, s}^{\top} \cdot \mathbf{x}_{s,(-h, t)} \cdot R_{\sigma, s} + b_{\sigma, s} ),
\end{gather}
where $L_{\mu, s}$, $L_{\sigma, s} \in \mathbb{R}^{d_s}$ and $R_{\mu, s}$, $R_{\sigma, s} \in \mathbb{R}^{h}$. $b_{\mu,s}$, while $b_{\sigma,s} \in \mathbb{R}$ are bias terms.
Note that the above parameters are data source specific and then the trainable set of parameters is denoted by $\theta_s \triangleq \{ L_{\mu, s}, L_{\sigma, s}, R_{\mu, s}, R_{\sigma,s}, b_{\mu,s}, b_{\sigma,s} \}$.

Based on the properties of log-normal distribution~\cite{mackay2003information, cohen1980estimation}, 
the mean and variance of the volume target on the original scale modeled by individual data sources can be derived from the mean and variance of log transformed volume as:
\begin{equation}\label{eq:m_s}
\mathbb{E}[{y}_{t} \, | \,  z_t = s, \mathbf{x}_{s, (-h,t)}, \theta_s] = \exp \{ \mu_{t,s} + \frac{1}{2} \cdot \sigma_{t,s}^2 \}  
\end{equation}
and
\begin{align}
\begin{split}\label{eq:v_s}
\mathbb{V}[{y}_{t} \, | \,  z_t = s, \mathbf{x}_{s, (-h,t)}, \theta_s] = & \exp\{\sigma_{t,s}^2 -1\} \cdot \exp\{2 \cdot \mu_{t,s} + \sigma_{t,s}^2\}    
\end{split}
\end{align}

Then, we define the probability distribution of the latent variable $z_t$ using a softmax function as follows:
\begin{equation}\label{eq:softmax}
\mathbb{P}_{\omega}(z_t = s \, | \, \mathbf{x}_{1, (-h,t)}, \cdots, \mathbf{x}_{S, (-h,t)}) \triangleq \frac{\exp(\, f_s(\mathbf{x}_{s, (-h, t)}) \, )}{ \exp( \, \sum_{k=1}^S f_k(\mathbf{x}_{k, (-h, t)}) \, )},
\end{equation}
where
\begin{equation}
f_s(\mathbf{x}_{s, (-h, t)} ) \triangleq L_{z, s}^{\top} \cdot \mathbf{x}_{s, (-h, t)} \cdot R_{z, s} + b_{z, s}
\end{equation}
$\omega \triangleq \{L_{z, s}, R_{z, s}, b_{z, s} \}_{s=1}^S$ denotes the set of trainable parameters regarding the latent variable $z_t$, i.e. $L_{z, s},R_{z, s} \in {\mathbb R}^{d_s}$ and $b_{z, s} \in {\mathbb R}$ is a bias term.

In the following, we will present how the distributions modeled by individual data sources and the latent variable will be used to learn the parameters in the training phase and to predict the mean and variance of the target volume in the predicting phase.

\subsection{Learning}

The learning process of TME is based on SGD optimization \cite{ruder2016overview, kingma2014adam}.
It is able to give rise to a set of parameter realizations for building the ensemble which has been proven to be an effective technique for enhancing the prediction accuracy as well as enabling uncertainty estimation in previous works
\cite{lakshminarayanan2017simple, maddox2019simple}.

We first briefly describe the training process of SGD optimization.
Denote the set of the parameters by $\Theta \triangleq \{\theta_1, \cdots, \theta_S, \omega\}$. 
The whole training dataset denoted by $\mathcal{D}$ is consisted of data instances, each of which is a pair of $y_t$ and $\{\mathbf{x}_{s, (-h,t)}\}_{1}^{S}$. 
$t$ is a time instant in the period $\{1, \cdots, T\}$.

Starting from a random initialized value of $\Theta$, in each iteration SGD samples a batch of training instances to update the model parameters as follows:
\begin{align}
\Theta(i) = \Theta(i-1) - \eta \nabla \mathcal{L}(\Theta(i-1); \mathcal{D}_i ),
\end{align}
where $i$ is the iteration step, $\Theta(i)$ represents the values of $\Theta$ at step $i$, i.e. a snapshot of $\Theta$.
$\eta$ is the learning rate, a tunable hyperparameter to control the magnitude of gradient update.
$\nabla \mathcal{L}(\Theta(i-1); \mathcal{D}_i)$ is the gradient of the loss function w.r.t. the model parameters given data batch $\mathcal{D}_i$ at iteration $i$.
The iteration stops when the loss converges with negligible variation.  
The model parameter snapshot at the last step or with the best validation performance will be taken as one realization of the model parameters.


The learning process of TME is to minimize the loss function defined by the negative log likelihood of the target volume as:
\begin{align}
\begin{split}
\mathcal{L}(\Theta;\mathcal{D}) \triangleq & - \sum^T_{t=1}\ln
\sum_{z_t = 1}^{S}
p_{\theta_s}(y_t | z_t = s, \mathbf{x}_{s, (-h,t)}) \cdot 
\mathbb{P}_{\omega}(z_t = s | \{ \mathbf{x}_{s, (-h,t)}\}_{1}^{S}) \\
& - \log p(\Theta),
\end{split}\label{eq:loss}
\end{align}
where the prior $p(\Theta)$ is viewed as a regularisation term.

By plugging the probability density function of log-normal distribution and Gaussian prior of $\Theta$ as a L2 regularisation, Eq.~\ref{eq:loss} is further expressed as: 
\begin{align}
\begin{split}
& - \sum^T_{t=1}\ln
\sum_{z_t = 1}^{S}
\frac{1}{y_t \sigma_{t, s} \sqrt{2\pi}} \exp\Big( -\frac{(\ln y_t  - \mu_{t,s})^2}{2\sigma^2_{t, s}} \Big) \cdot 
\mathbb{P}_{\omega}(z_t = s | \{ \mathbf{x}_{s, (-h,t)}\}_{1}^{S}) \\ 
&- \lambda \norm{\Theta}^2_2,
\end{split}\label{eq:loss_lognorm}
\end{align}
where $\lambda$ is the hyper-parameter of regularization strength.

In the SGD optimization process, different initialization of $\Theta$ leads to distinct parameter iterate trajectories.
Recent studies show that the ensemble of independently initialized and trained model parameters empirically often provide comparable performance on uncertainty quantification w.r.t. sampling and variational inference based methods~\cite{lakshminarayanan2017simple, snoek2019can, maddox2019simple}.
Our ensemble construction follows this idea by taking a set of model parameter realizations from different training trajectories, and each parameter realization is denoted by $\Theta_m = \{ \theta_{m,1}, \cdots, \theta_{m,S}, \omega_m \}$.
For more algorithmic details about the collecting procedure of these parameter realizations, please refer to the appendix.

\subsection{Prediction}\label{sec:infer}
In this part, we present the predicting process given model parameter realizations $\{ \Theta_m \}_1^M$.
We focus on two types of prediction quantities, i.e. mean and variance. 

Specifically, the predictive mean of the target in the original scale is defined as:
\begin{align}\label{eq:mean}
\mathbb{E}[{y}_{t} | \{\mathbf{x}_{s, (-h,t)}\}_{s=1}^S, \mathcal{D} ] \approx \frac{1}{M}\sum_{m=1}^M \mathbb{E}[{y}_t \, | \, \{\mathbf{x}_{s, (-h,t)}\}_{s=1}^S, \Theta_m ],
\end{align}
where $\mathbb{E}[{y}_t | \{\mathbf{x}_{s, (-h,t)}\}_{s=1}^S \, , \Theta_m ]$ is the conditional mean given one realization $\Theta_m$.
In TME, it is a weighted sum of the predictions by individual data sources as:
\begin{align}\label{eq:mean_sample}
\begin{split}
\mathbb{E}[{y}_t | \{\mathbf{x}_{s, (-h,t)}\}_{s=1}^S, \Theta_m ] = \sum_{s=1}^S & \mathbb{P}_{\omega_m}(z_{t} = s| \{\mathbf{x}_{s, (-h,t)}\}_{s=1}^S ) \\
& \cdot \mathbb{E}[{y}_{t} | z_t=s,\mathbf{x}_{s, (-h,t)}, \theta_{m,s}]
\end{split}
\end{align}

Apart from the predictive mean, the predictive variance is of great interest as well, since it helps to quantify the uncertainty on the prediction, thereby facilitating the downstream decision making based on the volume predictions.

Given the predictive mean, each data source's mean and variance defined in Eq.~\ref{eq:m_s}, \ref{eq:v_s} and \ref{eq:mean}, the predictive variance of the target volume is derived as:
\begin{align}\label{eq:var}
\begin{split}
&\mathbb{V}({y}_{t} | \{\mathbf{x}_{s, (-h,t)}\}_{s=1}^S, \mathcal{D}) \approx \int_y   y^2 p({y} | \{\mathbf{x}_{s, (-h,t)}\}_s, \mathcal{D} ) dy - \mathbb{E}^2[{y}_{t} | \{\mathbf{x}_{s, (-h,t)}\}_{s=1}^S, \mathcal{D} ] \\
& = \underbrace{\frac{1}{M} \sum_{m = 1}^M \sum_{s = 1}^S \mathbb{P}_{\omega_m}(z_t = s | \cdot)\mathbb{V}(y_t | z_t = s, \mathbf{x}_{s, (-h,t)}, \theta_{m,s})}_{\text{Aleatoric Uncertainty}} + \\
& \underbrace{ \frac{1}{M}\sum_{m = 1}^M \sum_{s = 1}^S \mathbb{P}_{\omega_m}(z_t = s | \cdot)
\mathbb{E}^2[y_t | z_t = s, \mathbf{x}_{s, (-h,t)}, \theta_{m,s}) - \mathbb{E}^{2}[y_t | \{\mathbf{x}_{s, (-h,t)}\}_{s=1}^S, \mathcal{D}],}_{\text{Epistemic Uncertainty}}
\end{split}
\end{align}
where for clarity of the formula $\mathbb{P}_{\omega_m}(z_{t} = s| \{\mathbf{x}_{s, (-h,t)}\}_{s=1}^S )$ is simplified to $\mathbb{P}_{\omega_m}(z_t = s | \cdot)$.

Eq.~\ref{eq:var} shows that the predictive variance is composed of two types of uncertainties, aleatoric and epistemic uncertainty.
The aleatoric part in Eq.~\ref{eq:var} stems from variance induced from multi-source data. 
It captures the noise inherent to the target which could depend on $\mathbf{x}_{s, (-h, t)}$.
As a comparison, the classical aleatoric uncertainty (or volatility) estimation model is typically used to estimate the uncertainty solely with the target time series or external features as a whole. 
It has no mechanism to capture the evolving relevance of multi-source data to the aleatoric uncertainty of the target.  
The epistemic uncertainty part in Eq.~\ref{eq:var} accounts for uncertainty in the model parameters i.e. uncertainty which captures our ignorance about which model generated our collected data.
It is also referred to as the model uncertainty. 

\section{Experiments}\label{sec:exp}

In this section, we report the overall experimental evaluation.
More detailed results are in the appendix section.

\subsection{Data and metrics}

\noindent
\textbf{Data}: 
We collected the limit order book and transaction data respectively from two exchanges, Bitfinex and Bitstamp and extracted features defined in Sec.~\ref{sec:data} from the order book and transactions of each exchange for the period from May 31, 2018 9:55pm (UTC) until September 30, 2018 9:59pm (UTC).
Then, for each exchange, we build three datasets of different prediction horizons, i.e. 1min, 5min, 10min, for training and evaluating models.
Depending on the prediction horizon, each instance in the dataset contains a target volume and the time-lagged features from order book and transactions of two exchanges.
In particular, for Bitfinex, the sizes of datasets for 1min, 5min, 10min are respectively $171727$, $34346$ and $17171$.
For Bitstamp, the sizes of datasets are respectively $168743$, $33749$ and $16873$.
In all the experiments, data instances are time ordered and we use the first $70\%$ of points for training, the next $10\%$ for validation, and the last $20\%$ of points for out-of-sample testing.
Note that all the metrics are evaluated on the out of sample testing data.

For modeling the log-normal distribution, the baseline models and TME need to perform the log-transformation on $y_t$ and thus in the data pre-processing step, we have filtered out the data instances with zero trading volume. 
Empirically, we found out that these zero-volume data instances account for less than $2.25\%$ and $3.98\%$ of the entire dataset for Bitfinex and Bitstamp markets respectively.


For the baseline methods not differentiating the data source of features, each target volume has a feature vector built by concatenating all features from order book and transactions of two markets into one vector.
For our TME, each target volume has four groups of features respectively corresponding to the order book and transaction data of two markets.

\noindent
\textbf{Metrics}: 
Note that baseline models and TME are trained to learn the log-normal distribution of the deseasonalized volume, however, the following metrics are evaluated on the original scale of the volume, because our aim is to quantify the performance on the real volume scale, that is of more interest for practical purposes.

In the following definitions, $\hat{v}_t$ corresponds to the predictive mean of the raw volume.
$\overline{T}$ is the number of data instances in the testing dataset.
$\hat{v}_t$ is derived from the predictive mean of $y_t$ multiplied by $a_{I(t)}$, according to the definition of $y_t$ in Sec.~\ref{sec:model}.
For baseline models, the predictive mean of $y_t$ is derived from the mean of logarithmic transformed $y_t$ via Eq.~\ref{eq:m_s} used by individual sources in TME.
The predictive mean of $y_t$ by TME is shown in Eq.\ref{eq:mean}. 
The considered performance metrics are:

\textit{RMSE:} is the root mean square error as $\text{RMSE}=\sqrt{\frac{1}{\overline{T}} \sum_{t=1}^{\overline{T}}(v_t-\hat{v}_t)^2}$.

\textit{MAE:} is mean absolute error as $\text{MAE}=\frac{1}{\overline{T}} \sum_{t=1}^{\overline{T}} |v_t-\hat{v}_t|$.

\textit{NNLL:} is the predictive Negative Log-Likelihood of testing instances normalized by the total number of testing points.
For the target variable volume, the likelihood of the real $v_t$ is that of the corresponding intraday-pattern free $y_t$ scaled by $\frac{1}{a_{I(t)}}$, according to the definition in Sec.\ref{sec:model}.
For TME, the likelihood of $y_t$ is calculated, as shown in Eq.~\ref{eq:loss_lognorm}, based on the mixture of log-normal distributions taking into account both predictive mean and variance.
For baseline models, the likelihood of $y_t$ can be straightforwardly derived from the predictive mean and variance of $\ln y_t$ ~\cite{mackay2003information, cohen1980estimation}.

\textit{IW:} interval width is meant to evaluate the uncertainty around the point prediction, i.e. predictive mean, in probabilistic forecasting.
The ideal model is expected to provide tight intervals, which imply the model is confident in the prediction.
In this paper, our baseline models and TME involve unimodal and multimodal distributions, i.e. log-normal and mixture of log-normal.
As a result, we choose to measure the interval width simply using the variance~\cite{kuleshov2018accurate}. Specifically, IW is defined as the averaged standard deviation of the testing instances. 
Note that similar to the scaling operation for getting $\hat{v}_t$, the standard deviation is obtained via firstly scaling the predictive variance of $y_t$ by $a_{I(t)}^2$. 


\subsection{Baseline models and TME setup}

As mentioned above, we benchmark the performance of TME against two econometric and one Machine Learning models. We present them below.

\noindent
\textbf{ARMA-GARCH} is the autoregressive moving average model (ARMA) plus generalized autoregressive conditional heteroskedasticity (GARCH) model~\cite{brownlees2010intra, satish2014predicting, chen2016forecasting}.
It is aimed to respectively capture the conditional mean and conditional variance of the logarithmic volume.
Then, the predictive mean and variance of the logarithmic volume are transformed to the original scale of the volume for evaluation.

The number of  autoregressive and moving average lags in ARMA are selected in the range from 1 to 10, by minimization of Akaike Information Criterion, while the orders of lag variances and lag residual errors in GARCH are found to affect the performance negligibly and thus are both fixed to one~\cite{Hansen2005}. The residual diagnostics for ARMA-GARCH models is given in the appendix Figure \ref{fig:acf_model_residuals_market1} and Figure \ref{fig:acf_model_residuals_market2}.

In Table~\ref{tab:garch_test} we report the estimated GARCH parameters in the ARMA-GARCH model on the log-volume, together with the standard errors and the p-value. 
All parameters are statistically different from zero at all time scales, indicating  the significant existence of heteroskedasticity.

\begin{table*}[!htbp]
\centering
 \caption{Statistics of GARCH(1,1) parameters ($\omega$, $\alpha$, $\beta$) on log-volume residuals, trained with ARMA($p,q$)-GARCH(1,1) model for different markets and prediction horizons. (*) indicate p-values $<10^{-5}$ for estimated parameters. Parameters $p,q$, were selected by minimizing AIC and are reported in Tables 2-4. }
\begin{tabular}{|c|c|c|c|c|c|c|}
  \hline
  \textbf{BITFINEX} & $\omega$ & Std.Err($\omega$)  &  $\alpha$ & Std.Err($\alpha$)  & $\beta$ & Std.Err($\beta$) \\ 
  \hline 
1 min  & 0.0177* & 0.0011 & 0.0259* & 0.0008  & 0.9677* & 0.0000  \\
5 min  & 0.0119* & 0.0015 & 0.0218* & 0.0017  & 0.9663* & 0.0000  \\
10 min & 0.0062* & 0.0011 & 0.0152* & 0.0018 & 0.9762* & 0.0000 \\
\hline
\textbf{BITSTAMP}& $\omega$ & Std.Err($\omega$)  &  $\alpha$ & Std.Err($\alpha$)  & $\beta$ & Std.Err($\beta$) \\ 
  \hline 
1 min  & 0.0112* & 0.0007 & 0.0203* & 0.0006 & 0.9759* & 0.0000 \\
5 min  & 0.0175* & 0.0023 & 0.0277* & 0.0022 & 0.9561* & 0.0000 \\
10 min & 0.0262* & 0.0056 & 0.0291* & 0.0042 & 0.9387* & 0.0000\\
  \hline
\end{tabular}
\label{tab:garch_test}
\end{table*}


\noindent
\textbf{ARMAX-GARCH} is the variant of ARMA-GARCH by adding external feature terms.
In our scenario, external features are obtained by concatenating all features from order book and transaction data of two exchanges into one feature vector. 
The hyper-parameters in ARMAX-GARCH are selected in the same way as for ARMA-GARCH.

\noindent
\textbf{GBM} is the gradient boosting machine \cite{friedman2001greedy}.
It has been empirically proven to be highly effective in predictive tasks across different machine learning challenges \cite{gulin2011winning,taieb2014gradient} and more recently in finance\cite{zhou2015evolution, sun2018novel}. 
The feature vector fed into GBM is also the concatenation of features from order book and transaction data of two markets. 
The hyper-parameters~\cite{scikit-learn} of GBM are selected by random search in the ranges: number of trees in $[100, \cdots, 1000]$, max number of features used by individual trees in $[1.0, \cdots, 0.1]$, minimum number of samples of the leaf nodes in $[2, \cdots, 9]$, maximum depth of individual trees in $[4, \cdots, 9]$, and the learning rate in $[0.05, \cdots, 0.005]$.

\noindent
\textbf{TME} is implemented by TensorFlow~\footnote{https://github.com/tensorflow/tensorflow}.
The hyper-parameters tuned by random search are mainly the learning rate in the range $[0.0001, \cdots, 0.001]$,
batch size in $[10, \cdots, 300]$, and the l2 regularization term in $[0.1, \cdots, 5.0]$.
The number of model parameter realizations for building the ensemble is set to 20. 
Beyond this number, we found no significant performance improvement.

\subsection{Results}

In this section, we present the results on the predictions of volume in both markets at different time scales.  Tables~\ref{tab:1min},~\ref{tab:5min}, and ~\ref{tab:10min} show the error metrics on the testing data for the two markets and the four models.
We observe that in all cases the smallest RMSE is achieved by TME while the smallest MAE is achieved by GBM. Concerning NNLL, in Bitfinex TME outperforms the other models for 1-min and 5-min cases, while for Bitstamp markets econometirc model have (slightly) lower values. Finally, the smallest IW is always achieved by TME.

The fact that GBM has superior performance on MAE is somewhat expected since GBM has been trained to minimize point prediction, while ARMA-GARCH, ARMAX-GARCH and TME were trained with a maximum likelihood objective.

By comparing RMSE and MAE in both markets, we observe that for ARMA-GARCH model in Bitfinex, external features and extra information from Bitstamp are lowering MAE and RMSE errors (except on 1min interval on Bitfinex market).
In Bitstamp market, for ARMAX-GARCH model external features only help on 1-min interval prediction. This phenomenon could indicate the one-directional information relevance across two markets.
However, due to the data source specific components and temporal adaptive weighting schema, our TME is able to yield more accurate prediction consistently, compared to ARMA-GARCH.

As for NNLL, similar pattern is observed in ARMA-GARCH family, i.e. additional features impair the NNLL performance instead, while TME retains comparable performance.
More importantly, TME has much lower IW.
Together with the lower RMSE, it implies that when TME predicts the mean closer to the observation, the predictive uncertainty is also lower.
Remind that GBM does not provide probabilistic predictions. Overall we find that TME outperforms quite often the baseline benchmarks, by providing smaller RMSE and tighter intervals.


\begin{table}[!htbp]
\begin{center}
  \caption{Results of 1-min volume prediction.
  The arrow symbols indicate the direction of the metrics for better models.
  GBM model is not providing uncertainty and NNLL and IW results are not available (NA). Results in bold, indicate the minimum errors among models.}
  \begin{tabular}{|c|c|c|c|c|c|c|}
    \hline
    \textbf{BITFINEX MARKET} & RMSE $\downarrow$ & MAE $\downarrow$ & NNLL $\downarrow$ & IW $\downarrow$\\
    \hline
ARMA(5,5)-GARCH(1,1)  & 24.547          & 14.227         & 2.660          & 117.348 \\
ARMAX(5,5)-GARCH(1,1) & 24.629          & 14.189         & 2.664          & 117.084 \\
GBM                   & 21.026          & \textbf{7.978} & NA             & NA \\
TME                   & \textbf{20.142} & 10.204         & \textbf{2.654} & \textbf{44.344} \\
    \hline
    \hline
    \textbf{BITSTAMP MARKET} & RMSE $\downarrow$ & MAE $\downarrow$ & NNLL $\downarrow$ & IW $\downarrow$\\
    \hline
ARMA(3,3)-GARCH(1,1) & 14.587            & 7.688           & \textbf{1.719} & 97.618\\
ARMAX(3,3)-GARCH(1,1) & 14.292           & 7.487           & 1.719          & 93.943\\
GBM                   & 11.740           & \textbf{3.515}  & NA             & NA\\
TME                   & \textbf{11.378}  & 4.299           & 1.720          & \textbf{10.295} \\
    \hline
\end{tabular}\label{tab:1min}
\end{center}
\end{table}

\begin{table}[!htbp]
\centering
\caption{Results of 5-min volume prediction.
The arrow symbols indicate the direction of the metrics for better models. GBM model is not providing uncertainty and NNLL and IW results are not available (NA). Results in bold, indicate the minimum errors among models.}
\begin{tabular}{|c|c|c|c|c|c|c|}
\hline
\textbf{BITFINEX MARKET} & RMSE $\downarrow$ & MAE $\downarrow$ & NNLL $\downarrow$ & IW $\downarrow$\\
\hline
ARMA(3,3)-GARCH(1,1)  & 64.999          & 39.909          & 4.642          & 82.158 \\
ARMAX(3,3)-GARCH(1,1) & 64.456          & 39.150          & 4.641          & 79.852 \\
GBM                   & 64.964          & \textbf{32.888} & NA             & NA \\
TME                   & \textbf{63.855} & 39.527          & \textbf{4.636} & \textbf{68.748} \\
\hline
\hline
\textbf{BITSTAMP MARKET} & RMSE $\downarrow$ & MAE $\downarrow$ & NNLL $\downarrow$ & IW $\downarrow$ \\
\hline
ARMA(5,4)-GARCH(1,1)  & 38.300          & 17.606          & \textbf{3.732} & 34.797\\
ARMAX(5,4)-GARCH(1,1) & 40.273          & 18.887          & 3.766          & 38.250\\
GBM                   & 39.196          & \textbf{14.714} & NA             & NA \\
TME                   & \textbf{38.223} & 17.287          & 3.765          & \textbf{29.148} \\
\hline
\end{tabular}\label{tab:5min}
\end{table}

\begin{table}[!htbp]
\begin{center}
  \caption{Results of 10-min volume prediction.
  The arrow symbols in the first line indicate the direction of the metrics for better models. GBM model is not providing uncertainty and NNLL and IW results are not available (NA)}
  \begin{tabular}{|c|c|c|c|c|c|c|c|}
    \hline
    \textbf{BITFINEX MARKET} & RMSE $\downarrow$ & MAE $\downarrow$ & NNLL $\downarrow$ & IW $\downarrow$ \\
    \hline
ARMA(3,2)-GARCH(1,1)  & 112.872          & 72.505           & 5.409          & 124.890\\
ARMAX(3,2)-GARCH(1,1) & 111.987          & 71.149           & \textbf{5.373} & 121.346\\
GBM                   & 110.197          & \textbf{61.506}  & NA             & NA\\
TME                   & \textbf{109.878} & 68.382           & 5.386          & \textbf{114.195}\\
    \hline
    \hline
    \textbf{BITSTAMP MARKET} & RMSE $\downarrow$ & MAE $\downarrow$ & NNLL $\downarrow$ & IW $\downarrow$ \\
    \hline
ARMA(3,2)-GARCH(1,1)  & 66.486          & 31.942          & \textbf{4.452} & 51.228\\
ARMAX(3,2)-GARCH(1,1) & 68.067          & 32.795          & 4.457          & 52.780\\
GBM                   & 67.128          & \textbf{27.719} & NA             & NA \\
TME                   & \textbf{66.234} & 31.460          & 4.507          & \textbf{49.972} \\
    \hline
\end{tabular}\label{tab:10min}
\end{center}
\end{table}

\begin{figure}[!htbp]
\centering
\includegraphics[width=0.98\textwidth]{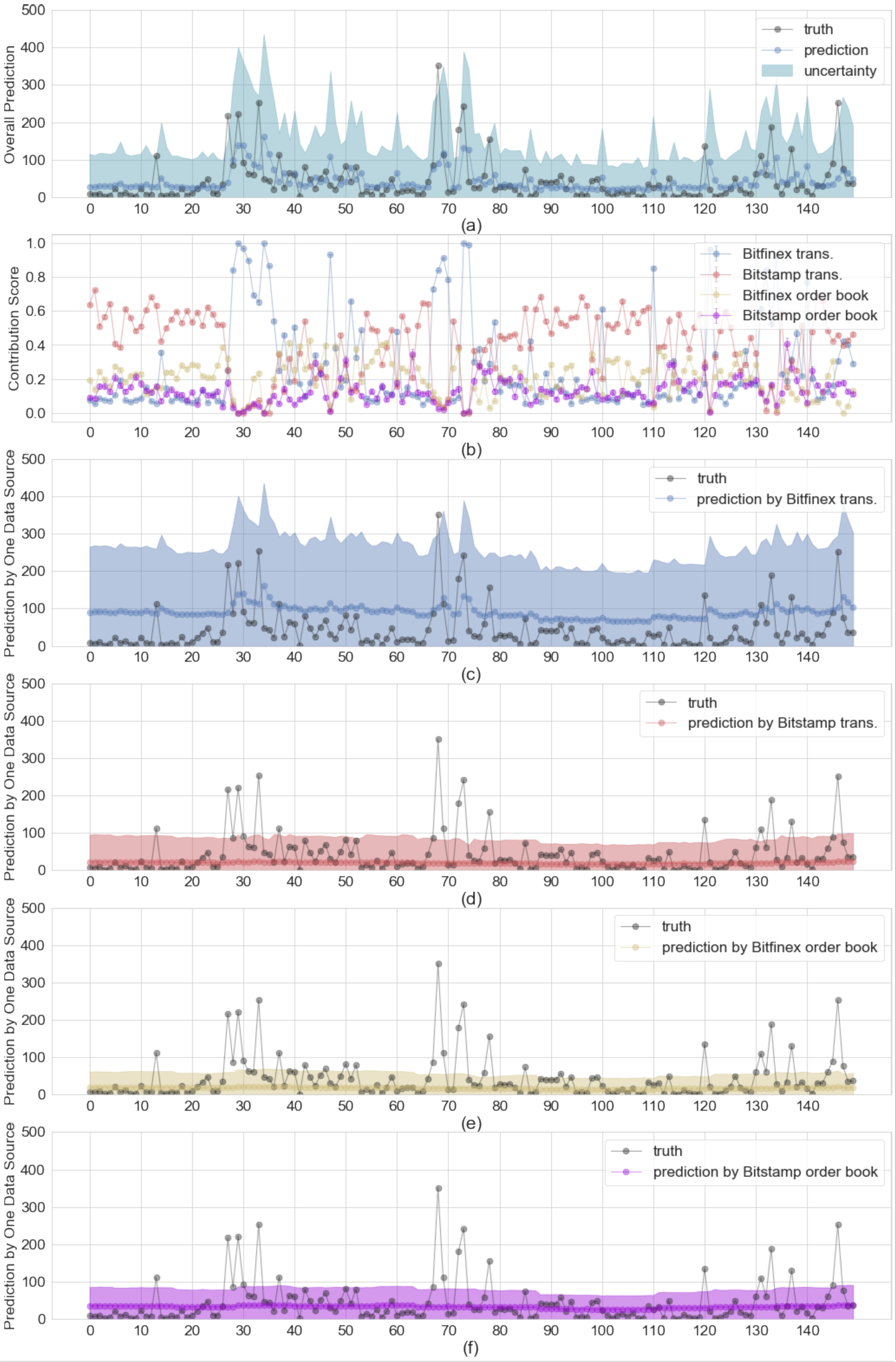}
\caption{Visualization of TME in a sample period of Bitfinex for 5-min volume predicting. 
Panel(a): Predictive mean $\pm$ two times standard deviation (left-truncated at zero) and the volume observations.
Panel(b): Data source contribution scores (i.e. average of latent variable probabilities) over time.
Panel(c)-(f): Each data source's predictive mean $\pm$ two times standard deviation (left-truncated at zero). 
The color of each source's plot corresponds to that of the contribution score in Panel(b).
(best viewed in colors)
}
\label{fig:vis_bitfinex5min}
\end{figure}

\begin{figure}[!htbp]
\centering
\includegraphics[width=0.98\textwidth]{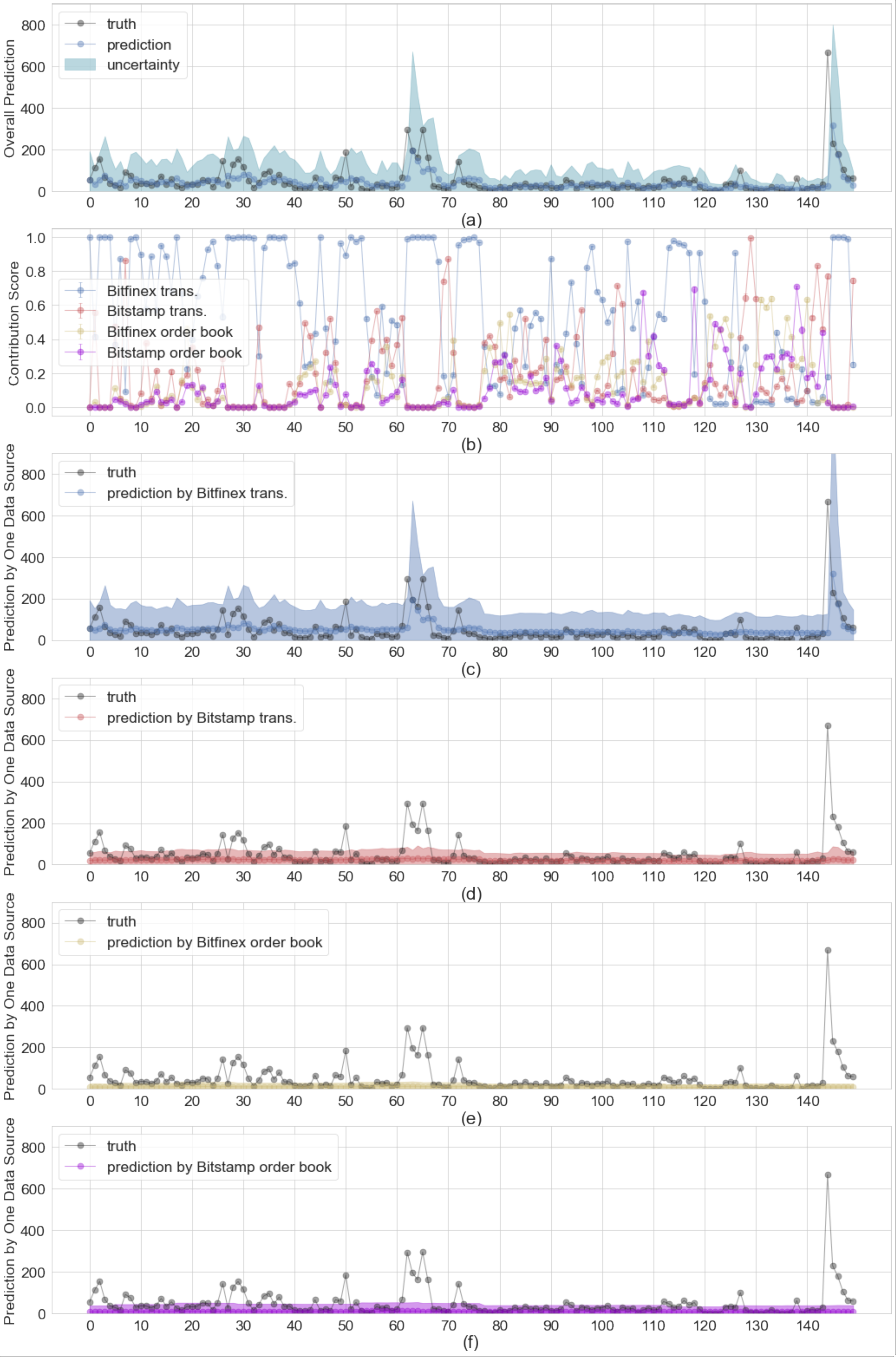}
\caption{Visualization of TME in a sample period of Bitstamp for 5-min volume predicting.
Panel(a): Predictive mean $\pm$ two times standard deviation (left-truncated at zero) and the volume observations.
Panel(b): Data source contribution scores (i.e. average of latent variable probabilities) over time.
Panel(c)-(f): Each data source's predictive mean $\pm$ two times standard deviation (left-truncated at zero). 
The color of each source's plot corresponds to that of the contribution score in Panel(b).
(best viewed in colors)
}
\label{fig:vis_bitstamp5min}
\end{figure}

In order to disentangle the different contributions to TME predictions and, at the same time, understanding why choosing different predicting features at each time step is important, we present in Fig.~\ref{fig:vis_bitfinex5min} and Fig.~\ref{fig:vis_bitstamp5min} the predictive volume and uncertainty of 5-min volume prediction in a sample time period of the testing data for the two markets.

Panel (a) shows how well the TME (pointwise and interval) predictions  follows the actual data. The TME is able to quantify at each time step the contribution of each source to the target forecasting. 
In Panel(b) we show the dynamical contribution scores of the four sources.
These scores are simply the average of the probability of the latent variable value corresponding to each data source, i.e. $z_t=s$.
We notice that the relative contributions varies with time and we observe that the external order book source from the less liquid market (Bitstamp) does not contribute much to predictions. 
On the contrary, in Panel (b) of Fig.~\ref{fig:vis_bitstamp5min}, where the data for Bitstamp are shown, external order book and external transaction features from the more liquid market (Bitfinex) play a more dominant role.
Then, when we further look at the predictions by individual data sources in TME, see Panel(c)-(f), what we observe is in line with the pattern of contribution scores in Panel(b).
For instance, the component model in TME responsible for the data source from more liquid Bitfinex captures more uncertainty, thereby being given high contribution scores in the volatile period. The addition of the features from the other markets allows to shrink the uncertainty intervals. We have also repeated these experiments for 1-min and 10-min volume prediction. 
The results are collected in the figures and tables in the appendix section. 

Finally, we discuss how well the different predictors perform conditionally to the volume size. To this end, we compute for each model, market, and time interval the RMSE and MAE conditional to the volume quartile. However, finding the suitable metrics to compare forecasting performance of a model across different quartiles is a subtle issue. Take for example a linear model for the log-volume. When considering it as a model for (linear) volume, it is clear that the additive noise in the log-volume becomes multiplicative in linear volume. Thus the RMSE conditional to volume becomes proportional to (or increasing as a power-law of) the volume and therefore one expects to see RMSE for high quartiles to be larger than the one for bottom quartiles. This is exactly what we observe in Table \ref{tab:5minQ1Q4} for 5 min horizons (for 1 and 10 min, see the appendix) when looking at RMSE and MAE. 

In order to take into account this statistical effect, essentially caused by the choice of modeling log-volume and presenting eroor metrics for linear volume, in Table \ref{tab:5minQ1Q4} we show also {\it relative} error metrics, namely relative root mean squared error (RelRMSE) and mean absolute percentage error (MAPE) conditioned to volume quartile. 
(RelRMSE is defined as $\sqrt{\frac{1}{\overline{T}} \sum_{t=1}^{\overline{T}}( \frac{v_t-\hat{v}_t}{v_t})^2}$), and MAPE is defined as $\frac{1}{\overline{T}} \sum_{t=1}^{\overline{T}}|\frac{v_t-\hat{v}_t}{v_t}|$.)
In this case the scenario changes completely. First, across all models the relative error is smaller for top volume quartiles and larger for bottom quartiles. Especially relative errors for the lowest quartile are large, likely because small volumes at the denominator create large fluctuations. Second, and more important, TME outperforms mostly in Q4, while GBM is superior in Q1, Q2, and Q£ frequently. The difference between models in Q4 is somewhat smaller, while in Q1-Q3 the out-performance of machine learning methods is up to a factor $2$ with respect to econometric methods. Thus also when considering large volumes and considering relative errors, TME and GBM provide more accurate predictions. 

\begin{table*}[htbp]
\centering
 \caption{5-min volume prediction errors conditional on the quartile of the true volume values.
  Measures: RMSE Qx -- root mean squared error conditioned on x-th quantile, RelRMSE Qx -- relative root mean squared error conditioned on x-th quantile, MAE Qx -- mean average error conditioned on x-th quantile, MAPE Qx -- mean absolute percentage error conditioned on x-th quantile.}
 \begin{tabular}{|c|c|c|c|c|}
 \hline
 \textbf{BITFINEX MARKET} & RMSE Q1 & RMSE Q2 & RMSE Q3 & RMSE Q4 \\ 
 \hline 
 ARMA-GARCH  & 32.862          & 38.594          & 43.073          & \textbf{111.660} \\
 ARMAX-GARCH & 31.939          & 36.887          & 41.005          & 112.020 \\
 GBM         & \textbf{17.632} & \textbf{17.332} & \textbf{22.056} & 125.604 \\
 TME         & 32.200          & 32.976          & 35.928          & 112.280 \\
 \hline
 & RelRMSE Q1 & RelRMSE Q2 & RelRMSE Q3 & RelRMSE Q4 \\ 
 \hline
 ARMA-GARCH  & 19.417          & 2.769          & 1.299          & 0.609\\
 ARMAX-GARCH & 18.845          & 2.657          & 1.243          & 0.594\\
 GBM         & \textbf{12.672} & \textbf{1.268} & \textbf{0.615} & 0.623\\
 TME         & 26.235          & 2.378          & 1.098          & \textbf{0.578}\\
 \hline 
 & MAE Q1 & MAE Q2 & MAE Q3 & MAE Q4 \\
 \hline
 ARMA-GARCH & 26.781 & 28.635 & 29.622 & 74.576\\
 ARMAX-GARCH & 26.145          & 27.625          & 28.190          & 74.612 \\
 GBM         & \textbf{14.719} & \textbf{11.985} & \textbf{16.674} & 88.122 \\
 TME         & 26.317          & 28.784          & 23.338          & \textbf{74.484} \\
 \hline
 & MAPE Q1 & MAPE Q2 & MAPE Q3 & MAPE Q4 \\
 \hline
 ARMA-GARCH  & 9.506          & 2.003          & 0.871          & 0.497\\
 ARMAX-GARCH & 9.272          & 1.935          & 0.831          & \textbf{0.493}\\
 GBM         & \textbf{5.717} & \textbf{0.846} & \textbf{0.460} & 0.570\\
 TME         & 11.424         & 1.823          & 0.690          & 0.503\\
 \hline
 \hline
 \textbf{BITSTAMP MARKET} & RMSE Q1 & RMSE Q2 & RMSE Q3 & RMSE Q4 \\ 
 \hline
ARMA-GARCH  & 13.465 & 15.304 & 16.199 & \textbf{72.046}\\
 ARMAX-GARCH & 14.2 & 16.835 & 20.0009 & 74.8532\\
 GBM         & \textbf{6.432}  & \textbf{6.679} & \textbf{8.732} & 77.317 \\
 TME         & 13.617          & 14.492         & 14.583         & 73.778 \\
 \hline
  & RelRMSE Q1 & RelRMSE Q2 & RelRMSE Q3 & RelRMSE Q4 \\ 
 \hline
 ARMA-GARCH  & 22.171          & 2.563          & 1.175          & 0.671\\
 ARMAX-GARCH & 24.259          & 2.784          & 1.429          & 0.889\\
 GBM         & \textbf{12.578} & \textbf{1.154} & \textbf{0.596} & 0.658\\
 TME         & 29.289          & 2.429          & 1.084          & \textbf{0.586}\\
 \hline
 & MAE Q1 & MAE Q2 & MAE Q3 & MAE Q4 \\
 \hline
 ARMA-GARCH  & 10.668 & 11.648 & 11.020 & 37.094\\
 ARMAX-GARCH & 11.0917 & 12.275 & 12.1831 & 39.9903\\
 GBM         & \textbf{5.037} & \textbf{4.575} & \textbf{6.705} & 42.506\\
 TME         & 11.778         & 11.342         & 10.138         & \textbf{36.827}\\
 \hline
 & MAPE Q1 & MAPE Q2 & MAPE Q3 & MAPE Q4 \\
 \hline
 ARMA-GARCH & 11.049         & 1.910          & 0.792          & 0.523\\
 ARMA-GARCH & 11.656         & 2.007          & 0.873          & 0.603\\
 GBM        & \textbf{5.838} & \textbf{0.755} & \textbf{0.456} & 0.599\\
 TME        & 14.117         & 1.871          & 0.720          & \textbf{0.493}\\
 \hline
\end{tabular}
\label{tab:5minQ1Q4}
\end{table*}

\newpage

\section{Conclusion and discussion}\label{sec:conclusion}

In this paper, we analyzed the problem of predicting trading volume and its uncertainty in cryptocurrency exchange markets.
The main innovations proposed in this paper are (i) the use of transaction and order book data from different markets and (ii) the use of TME, a class of models able to identify at each time step the set of data locally more useful in predictions. 

By investigating data from BTC/USD exchange markets, we found that  time series models of the ARMA-GARCH family do provide fair basic predictions for volume and its uncertainty, but when external data (e.g. from order book and/or from other markets) are added, the prediction performance does not improve significantly. Our analysis suggests that this might be due to the fact that the contribution of this data to the prediction could be not constant over time, but depending on the "market state". The temporal mixture ensemble model is designed precisely to account for such a variability. Indeed we find that this method outperforms time series models both in point and in interval predictions of trading volume. Moreover, especially when compared to other machine learning methods, the temporal mixture approach is significantly more interpretable, allowing the inference of the dynamical contributions from different data sources as a core part of the learning procedure. This has important potential implications for decision making in economics and finance. 

Also when conditioning to volume quartile, TME and GBM outperform econometric methods especially in the first three quartiles. For large volumes, likely due to the presence of unexpected bursts of volume which are very challenging to forecast, the performances of the methods are more comparable. However by using relative RMSE and MAPE the forecasting errors for large volumes are small.


Finally, although the method has been proposed and tested for cryptocurrency volume in two specific exchanges, we argue that it can be successfully applied (in future work) to other cryptocurrencies and to more traditional financial assets. 



\section*{Acknowledgements}
This work has been funded by the European  Program scheme 'INFRAIA-01-2018-2019: Research and Innovation action', grant agreement \#871042 'SoBigData++: European Integrated Infrastructure for Social Mining and Big Data Analytics'. We would like also to thank two anonymous referees for their very detailed reports. Their suggestions contributed significantly to the improvement of the paper.

\bibliographystyle{IEEEtran}
\bibliography{reference} 

\begin{thebibliography}{10}
\providecommand{\url}[1]{#1}
\csname url@samestyle\endcsname
\providecommand{\newblock}{\relax}
\providecommand{\bibinfo}[2]{#2}
\providecommand{\BIBentrySTDinterwordspacing}{\spaceskip=0pt\relax}
\providecommand{\BIBentryALTinterwordstretchfactor}{4}
\providecommand{\BIBentryALTinterwordspacing}{\spaceskip=\fontdimen2\font plus
\BIBentryALTinterwordstretchfactor\fontdimen3\font minus
  \fontdimen4\font\relax}
\providecommand{\BIBforeignlanguage}[2]{{%
\expandafter\ifx\csname l@#1\endcsname\relax
\typeout{** WARNING: IEEEtran.bst: No hyphenation pattern has been}%
\typeout{** loaded for the language `#1'. Using the pattern for}%
\typeout{** the default language instead.}%
\else
\language=\csname l@#1\endcsname
\fi
#2}}
\providecommand{\BIBdecl}{\relax}
\BIBdecl

\bibitem{urquhart2016inefficiency}
A.~Urquhart, ``The inefficiency of bitcoin,'' \emph{Economics Letters}, vol.
  148, pp. 80--82, 2016.

\bibitem{Bolt2016}
W.~Bolt, ``On the value of virtual currencies,'' \emph{{SSRN} Electronic
  Journal}, 2016.

\bibitem{cheah2015speculative}
E.-T. Cheah and J.~Fry, ``Speculative bubbles in bitcoin markets? an empirical
  investigation into the fundamental value of bitcoin,'' \emph{Economics
  Letters}, vol. 130, pp. 32--36, 2015.

\bibitem{Chu2015}
J.~Chu, S.~Nadarajah, and S.~Chan, ``Statistical analysis of the exchange rate
  of bitcoin,'' \emph{PLOS ONE}, vol.~10, no.~7, pp. 1--27, 2015.

\bibitem{BouchaudBTC}
J.~Donier and J.-P. Bouchaud, ``Why do markets crash? bitcoin data offers
  unprecedented insights,'' \emph{PLOS ONE}, vol.~10, pp. 1--11, 2015.

\bibitem{ciaian2016economics}
P.~Ciaian, M.~Rajcaniova, and d.~Kancs, ``The economics of bitcoin price
  formation,'' \emph{Applied Economics}, vol.~48, pp. 1799--1815, 2016.

\bibitem{Ron2013BTC}
D.~Ron and A.~Shamir, ``Quantitative analysis of the full bitcoin transaction
  graph,'' in \emph{International Conference on Financial Cryptography and Data
  Security}.\hskip 1em plus 0.5em minus 0.4em\relax Springer, 2013, pp. 6--24.

\bibitem{jang2018empirical}
H.~Jang and J.~Lee, ``An empirical study on modeling and prediction of bitcoin
  prices with bayesian neural networks based on blockchain information,''
  \emph{IEEE Access}, vol.~6, pp. 5427--5437, 2018.

\bibitem{amjad2017trading}
M.~Amjad and D.~Shah, ``Trading bitcoin and online time series prediction,'' in
  \emph{NIPS 2016 Time Series Workshop}, 2017, pp. 1--15.

\bibitem{alessandretti2018machine}
L.~Alessandretti, A.~ElBahrawy, L.~M. Aiello, and A.~Baronchelli, ``Machine
  learning the cryptocurrency market,'' \emph{arXiv preprint arXiv:1805.08550},
  2018.

\bibitem{guo2018bitcoin}
T.~Guo, A.~Bifet, and N.~Antulov-Fantulin, ``Bitcoin volatility forecasting
  with a glimpse into buy and sell orders,'' in \emph{2018 IEEE International
  Conference on Data Mining (ICDM)}.\hskip 1em plus 0.5em minus 0.4em\relax
  IEEE, 2018, pp. 989--994.

\bibitem{Garcia2015}
D.~Garcia and F.~Schweitzer, ``Social signals and algorithmic trading of
  bitcoin,'' \emph{Royal Society Open Science}, vol.~2, no.~9, p. 150288, 2015.

\bibitem{wheatley2019bitcoin}
S.~Wheatley, D.~Sornette, T.~Huber, M.~Reppen, and R.~N. Gantner, ``Are bitcoin
  bubbles predictable? combining a generalized metcalfe’s law and the
  log-periodic power law singularity model,'' \emph{Royal Society open
  science}, vol.~6, no.~6, p. 180538, 2019.

\bibitem{gerlach2019dissection}
J.-C. Gerlach, G.~Demos, and D.~Sornette, ``Dissection of bitcoin’s
  multiscale bubble history from january 2012 to february 2018,'' \emph{Royal
  Society open science}, vol.~6, no.~7, p. 180643, 2019.

\bibitem{antulov2018inferring}
N.~Antulov-Fantulin, D.~Tolic, M.~Piskorec, Z.~Ce, and I.~Vodenska, ``Inferring
  short-term volatility indicators from the bitcoin blockchain,'' in
  \emph{International Conference on Complex Networks and their
  Applications}.\hskip 1em plus 0.5em minus 0.4em\relax Springer, 2018, pp.
  508--520.

\bibitem{KondorBTC}
D.~Kondor, I.~Csabai, J.~Szule, M.~Posfai, and G.~Vattay, ``Inferring the
  interplay between network structure and market effects in bitcoin,''
  \emph{New Journal of Physics}, vol.~16, p. 125003, 2014.

\bibitem{ElBahrawy2017}
A.~ElBahrawy, L.~Alessandretti, A.~Kandler, R.~Pastor-Satorras, and
  A.~Baronchelli, ``Evolutionary dynamics of the cryptocurrency market,''
  \emph{Royal Society Open Science}, vol.~4, no.~11, p. 170623, 2017.

\bibitem{Nakamoto2008}
\BIBentryALTinterwordspacing
S.~Nakamoto, ``Bitcoin: A peer-to-peer electronic cash system,'' 2008.
  [Online]. Available: \url{http://bitcoin.org/bitcoin.pdf}
\BIBentrySTDinterwordspacing

\bibitem{bos2014elliptic}
J.~W. Bos, J.~A. Halderman, N.~Heninger, J.~Moore, M.~Naehrig, and E.~Wustrow,
  ``Elliptic curve cryptography in practice,'' in \emph{International
  Conference on Financial Cryptography and Data Security}.\hskip 1em plus 0.5em
  minus 0.4em\relax Springer, 2014, pp. 157--175.

\bibitem{mayer2016ecdsa}
H.~Mayer, ``Ecdsa security in bitcoin and ethereum: a research survey,''
  \emph{CoinFaabrik, June}, vol.~28, p. 126, 2016.

\bibitem{jakobsson1999proofs}
M.~Jakobsson and A.~Juels, ``Proofs of work and bread pudding protocols,'' in
  \emph{Secure information networks}.\hskip 1em plus 0.5em minus 0.4em\relax
  Springer, 1999, pp. 258--272.

\bibitem{BAUMOHL2019363}
E.~Baumöhl, ``Are cryptocurrencies connected to forex? a quantile
  cross-spectral approach,'' \emph{Finance Research Letters}, vol.~29, pp. 363
  -- 372, 2019.

\bibitem{algo1}
A.~P. Chaboud, B.~Chiquoine, H.~E., and C.~Vega, ``Rise of the machines:
  Algorithmic trading in the foreign exchange market,'' \emph{The Journal of
  Finance}, vol.~69, no.~5, pp. 2045--2084, 2014.

\bibitem{algo2}
T.~Hendershott, C.~Jones, and A.~Menkveld, ``Does algorithmic trading improve
  liquidity?'' \emph{The Journal of Finance}, vol.~66, no.~1, pp. 1--33, 2011.

\bibitem{frei2015optimal}
C.~Frei and N.~Westray, ``Optimal execution of a vwap order: a stochastic
  control approach,'' \emph{Mathematical Finance}, vol.~25, no.~3, pp.
  612--639, 2015.

\bibitem{barzykin2019optimal}
A.~Barzykin and F.~Lillo, ``Optimal vwap execution under transient price
  impact,'' \emph{arXiv preprint arXiv:1901.02327}, 2019.

\bibitem{brownlees2010intra}
C.~T. Brownlees, F.~Cipollini, and G.~M. Gallo, ``Intra-daily volume modeling
  and prediction for algorithmic trading,'' \emph{Journal of Financial
  Econometrics}, vol.~9, no.~3, pp. 489--518, 2010.

\bibitem{satish2014predicting}
V.~Satish, A.~Saxena, and M.~Palmer, ``Predicting intraday trading volumeand
  volume percentages,'' \emph{The journal of trading}, vol.~9, no.~3, pp.
  15--25, 2014.

\bibitem{chen2016forecasting}
R.~Chen, Y.~Feng, and D.~Palomar, ``Forecasting intraday trading volume: A
  kalman filter approach,'' \emph{Available at SSRN 3101695}, 2016.

\bibitem{bialkowski2008improving}
J.~Bialkowski, S.~Darolles, and G.~Le~Fol, ``Improving vwap strategies: A
  dynamic volume approach,'' \emph{Journal of Banking \& Finance}, vol.~32,
  no.~9, pp. 1709--1722, 2008.

\bibitem{calvori2013go}
F.~Calvori, F.~Cipollini, and G.~M. Gallo, ``Go with the flow: A gas model for
  predicting intra-daily volume shares,'' \emph{Available at SSRN 2363483},
  2013.

\bibitem{kawakatsu2018direct}
H.~Kawakatsu, ``Direct multiperiod forecasting for algorithmic trading,''
  \emph{Journal of Forecasting}, vol.~37, no.~1, pp. 83--101, 2018.

\bibitem{andersen1996-MDH}
T.~G. Andersen, ``Return volatility and trading volume: An information flow
  interpretation of stochastic volatility,'' \emph{The Journal of Finance},
  vol.~51, no.~1, pp. 169--204, 1996.

\bibitem{Katsiampa2017}
P.~Katsiampa, ``Volatility estimation for bitcoin: A comparison of {GARCH}
  models,'' \emph{Economics Letters}, vol. 158, pp. 3--6, 2017.

\bibitem{balcilar2017can}
M.~Balcilar, E.~Bouri, R.~Gupta, and D.~Roubaud, ``Can volume predict bitcoin
  returns and volatility? a quantiles-based approach,'' \emph{Economic
  Modelling}, vol.~64, pp. 74--81, 2017.

\bibitem{hougan2019SEC}
M.~Hougan, H.~Kim, M.~Lerner, and B.~A. Management, ``Economic and non-economic
  trading in bitcoin: Exploring the real spot market for the world’s first
  digital commodity,'' \emph{Bitwise Asset Management}, 2019.

\bibitem{OB_rev_Porter2013}
M.~D. Gould, M.~A. Porter, S.~Williams, M.~McDonald, D.~J. Fenn, and S.~D.
  Howison, ``Limit order books,'' \emph{Quantitative Finance}, vol.~13, no.~11,
  pp. 1709--1742, 2013.

\bibitem{Lillo2016}
M.~Rambaldi, E.~Bacry, and F.~Lillo, ``The role of volume in order book
  dynamics: a multivariate hawkes process analysis,'' \emph{Quantitative
  Finance}, vol.~17, no.~7, pp. 999--1020, 2016.

\bibitem{andersen1997intraday}
T.~G. Andersen and T.~Bollerslev, ``Intraday periodicity and volatility
  persistence in financial markets,'' \emph{Journal of empirical finance},
  vol.~4, no. 2-3, pp. 115--158, 1997.

\bibitem{bauwens2008moments}
L.~Bauwens, F.~Galli, and P.~Giot, ``Moments of the log-acd model,''
  \emph{Quantitative and Qualitative Analysis in Social Sciences}, vol.~2, pp.
  1--28, 2008.

\bibitem{engle2002new}
R.~Engle, ``New frontiers for arch models,'' \emph{Journal of Applied
  Econometrics}, vol.~17, no.~5, pp. 425--446, 2002.

\bibitem{ComponentGARCH}
R.~F. Engle and M.~E. Sokalska, ``Forecasting intraday volatility in the us
  equity market. multiplicative component garch,'' \emph{Journal of Financial
  Econometrics}, vol.~10, no.~1, pp. 54--83, 2012.

\bibitem{PGARCH}
T.~Bollerslev and E.~Ghysels, ``Periodic autoregressive conditional
  heteroscedasticity,'' \emph{Journal of Business \& Economic Statistics},
  vol.~14, no.~2, pp. 139--151, 1996.

\bibitem{waterhouse1996bayesian}
S.~R. Waterhouse, D.~MacKay, and A.~J. Robinson, ``Bayesian methods for
  mixtures of experts,'' in \emph{Advances in neural information processing
  systems}, 1996, pp. 351--357.

\bibitem{yuksel2012twenty}
S.~E. Yuksel, J.~N. Wilson, and P.~D. Gader, ``Twenty years of mixture of
  experts,'' \emph{IEEE transactions on neural networks and learning systems},
  vol.~23, pp. 1177--1193, 2012.

\bibitem{wei2007dynamic}
X.~Wei, J.~Sun, and X.~Wang, ``Dynamic mixture models for multiple
  time-series.'' in \emph{IJCAI}, vol.~7, 2007, pp. 2909--2914.

\bibitem{bazzani2016recurrent}
L.~Bazzani, H.~Larochelle, and L.~Torresani, ``Recurrent mixture density
  network for spatiotemporal visual attention,'' \emph{arXiv preprint
  arXiv:1603.08199}, 2016.

\bibitem{guo2019exploring}
T.~Guo, T.~Lin, and N.~Antulov-Fantulin, ``Exploring interpretable lstm neural
  networks over multi-variable data,'' in \emph{International Conference on
  Machine Learning}, 2019, pp. 2494--2504.

\bibitem{schwab2019granger}
P.~Schwab, D.~Miladinovic, and W.~Karlen, ``Granger-causal attentive mixtures
  of experts: Learning important features with neural networks,'' in
  \emph{Proceedings of the AAAI Conference on Artificial Intelligence},
  vol.~33, 2019, pp. 4846--4853.

\bibitem{kurle2019multi}
R.~Kurle, S.~G{\"u}nnemann, and P.~van~der Smagt, ``Multi-source neural
  variational inference,'' in \emph{Proceedings of the AAAI Conference on
  Artificial Intelligence}, vol.~33, 2019, pp. 4114--4121.

\bibitem{lakshminarayanan2017simple}
B.~Lakshminarayanan, A.~Pritzel, and C.~Blundell, ``Simple and scalable
  predictive uncertainty estimation using deep ensembles,'' in \emph{Advances
  in neural information processing systems}, 2017, pp. 6402--6413.

\bibitem{maddox2019simple}
W.~J. Maddox, P.~Izmailov, T.~Garipov, D.~P. Vetrov, and A.~G. Wilson, ``A
  simple baseline for bayesian uncertainty in deep learning,'' in
  \emph{Advances in Neural Information Processing Systems}, 2019, pp.
  13\,132--13\,143.

\bibitem{snoek2019can}
J.~Snoek, Y.~Ovadia, E.~Fertig, B.~Lakshminarayanan, S.~Nowozin, D.~Sculley,
  J.~Dillon, J.~Ren, and Z.~Nado, ``Can you trust your model's uncertainty?
  evaluating predictive uncertainty under dataset shift,'' in \emph{Advances in
  Neural Information Processing Systems}, 2019, pp. 13\,969--13\,980.

\bibitem{mackay2003information}
D.~J. MacKay and D.~J. Mac~Kay, \emph{Information theory, inference and
  learning algorithms}.\hskip 1em plus 0.5em minus 0.4em\relax Cambridge
  university press, 2003.

\bibitem{cohen1980estimation}
A.~C. Cohen and B.~J. Whitten, ``Estimation in the three-parameter lognormal
  distribution,'' \emph{Journal of the American Statistical Association},
  vol.~75, no. 370, pp. 399--404, 1980.

\bibitem{ruder2016overview}
S.~Ruder, ``An overview of gradient descent optimization algorithms,''
  \emph{arXiv preprint arXiv:1609.04747}, 2016.

\bibitem{kingma2014adam}
D.~P. Kingma and J.~Ba, ``Adam: A method for stochastic optimization,''
  \emph{International Conference on Learning Representations}, 2015.

\bibitem{kuleshov2018accurate}
V.~Kuleshov, N.~Fenner, and S.~Ermon, ``Accurate uncertainties for deep
  learning using calibrated regression,'' \emph{arXiv preprint
  arXiv:1807.00263}, 2018.

\bibitem{Hansen2005}
P.~R. Hansen and A.~Lunde, ``A forecast comparison of volatility models: does
  anything beat a {GARCH}(1, 1)?'' \emph{Journal of Applied Econometrics},
  vol.~20, pp. 873--889, 2005.

\bibitem{friedman2001greedy}
J.~H. Friedman, ``Greedy function approximation: a gradient boosting machine,''
  \emph{Annals of statistics}, pp. 1189--1232, 2001.

\bibitem{gulin2011winning}
A.~Gulin, I.~Kuralenok, and D.~Pavlov, ``Winning the transfer learning track of
  yahoo!’s learning to rank challenge with yetirank,'' in \emph{Proceedings
  of the Learning to Rank Challenge}, 2011, pp. 63--76.

\bibitem{taieb2014gradient}
S.~B. Taieb and R.~J. Hyndman, ``A gradient boosting approach to the kaggle
  load forecasting competition,'' \emph{International journal of forecasting},
  vol.~30, no.~2, pp. 382--394, 2014.

\bibitem{zhou2015evolution}
N.~Zhou, W.~Cheng, Y.~Qin, and Z.~Yin, ``Evolution of high-frequency systematic
  trading: a performance-driven gradient boosting model,'' \emph{Quantitative
  Finance}, vol.~15, no.~8, pp. 1387--1403, 2015.

\bibitem{sun2018novel}
X.~Sun, M.~Liu, and Z.~Sima, ``A novel cryptocurrency price trend forecasting
  model based on lightgbm,'' \emph{Finance Research Letters}, 2018.

\bibitem{scikit-learn}
F.~Pedregosa, G.~Varoquaux, A.~Gramfort, V.~Michel, B.~Thirion, O.~Grisel,
  M.~Blondel, P.~Prettenhofer, R.~Weiss, V.~Dubourg, J.~Vanderplas, A.~Passos,
  D.~Cournapeau, M.~Brucher, M.~Perrot, and E.~Duchesnay, ``Scikit-learn:
  Machine learning in {P}ython,'' \emph{Journal of Machine Learning Research},
  vol.~12, pp. 2825--2830, 2011.

\bibitem{GARCH}
T.~Bollerslev, ``Generalized autoregressive conditional heteroskedasticity,''
  \emph{Journal of econometrics}, vol.~31, no.~3, pp. 307--327, 1986.

\bibitem{bentejac2020comparative}
C.~Bent{\'e}jac, A.~Cs{\"o}rg{\H{o}}, and G.~Martinez-Munoz, ``A comparative
  analysis of gradient boosting algorithms,'' \emph{Artificial Intelligence
  Review}, pp. 1--31, 2020.

\bibitem{chen2016xgboost}
T.~Chen and C.~Guestrin, ``Xgboost: A scalable tree boosting system,'' in
  \emph{Proceedings of the 22nd acm sigkdd international conference on
  knowledge discovery and data mining}, 2016, pp. 785--794.

\bibitem{lu2020randomized}
H.~Lu and R.~Mazumder, ``Randomized gradient boosting machine,'' \emph{SIAM
  Journal on Optimization}, vol.~30, no.~4, pp. 2780--2808, 2020.

\bibitem{mandt2017stochastic}
S.~Mandt, M.~D. Hoffman, and D.~M. Blei, ``Stochastic gradient descent as
  approximate bayesian inference,'' \emph{The Journal of Machine Learning
  Research}, vol.~18, no.~1, pp. 4873--4907, 2017.

\bibitem{gur2018gradient}
G.~Gur-Ari, D.~A. Roberts, and E.~Dyer, ``Gradient descent happens in a tiny
  subspace,'' \emph{arXiv preprint arXiv:1812.04754}, 2018.

\end{thebibliography}
\newpage

\section{Appendix}

\subsection{ARMAX-GARCH}

As mentioned, our benchmarks belong to the ARMAX-GARCH class with external regressors. 
More specifically, the volume process is modelled with the following:\\
\begin{equation}
    \Phi(L)(\ln(y_t) - \mu_t) = \Theta(L)\epsilon_t,
\end{equation}
where $\Phi(L)$, $\Theta(L)$ denote polynomials of the lag operator $L$. The time varying mean $\mu_t$ is modeled as
\begin{equation}
    \mu_t = \mu + \sum_{s=1}^{S} \sum_{j=1}^{d_s}\psi_{s,j} x_{s,t-1}(j)
\end{equation}
where $x_{s,t-1}(j)$ denotes the $j$-th feature from external feature vector $\mathbf{x}_{s,t-1}$ at time $t-1$ from source $s$. The total number of sources $S=4$, which includes transactions and limit order book data of the two markets.  
Since variance of volume might exhibit time clustering, we assume that the residuals $\epsilon_t$ are modelled by a GARCH process \cite{GARCH,brownlees2010intra, satish2014predicting, chen2016forecasting}:
\begin{equation}
\epsilon_t = \sigma_t e_t~~~~~~~ e_t \sim \mathcal{N}(0,1)
\end{equation}
 \begin{equation}
\sigma_t^{2} = \omega + \alpha\epsilon_{t-1}^{2}  + \beta\sigma_{t-1}^{2}
 \end{equation}

\subsection{Gradient boosting}
In the following, we summarize how GBM is the used in the context of volume predicting. 
For more details of GBM, we suggest referring to \cite{friedman2001greedy}.

At time $t$, the target label is $u_t=\ln{y_{t+1}}$, is the logarithm of deseasonalized volume at next time segment.
Gradient boosting approximates the target variable $u_t$ with a function $F(\mathbf{x}_{t})$ that has the following additive expansion (similar to other functional approximation methods like radial basis functions, neural networks, wavelets, etc.):
\begin{equation}
\hat{u_{t}}=F(\mathbf{x}_{t})=\sum_{m=0}^M \beta_m h(\mathbf{x}_{t};\mathbf{a}_m),
\end{equation}
where $\mathbf{x}_{t}$ denotes the feature vector, that is constructed as a concatenation from different sources\footnote{Note, that we have omitted the transpose operators in the next line, as the concatenation is simple operation and to avoid confusion with index of time.} $\mathbf{x}_{t}=(\mathbf{x}_{s=1, (-h,t)},\mathbf{x}_{s=2, (-h,t)},\mathbf{x}_{s=3, (-h,t)},\mathbf{x}_{s=4, (-h,t)})$.

For a given training sample $\{ u_{t}, \mathbf{x}_t \}_{t=1}^T$, 
our goal is to find a function $F^*(\mathbf{x})$ such that the expected value of loss function $\frac{1}{2}(u - F(\mathbf{x}))^2$ ({squared loss}) is minimized over the joint distribution of $\{ u, \mathbf{x} \}$
\begin{equation}
F^*(\mathbf{x}) = \argmin_{F(x)} \mathbb{E}_{u, \mathbf{x}} (u- F(\mathbf{x}))^2.
\end{equation}
Under the additive expansion $F(\mathbf{x})=\sum_{m=0}^M \beta_m h(\mathbf{x};\mathbf{a}_m)$ with parameterized functions $h(\mathbf{x};\mathbf{a}_m)$, we proceed 
by making the initial guess $F_0(\mathbf{x})=\argmin_{c} \sum_{t=1}^T (u_{t}- c)^2$ and then parameters are jointly fit in a forward incremental way $m=1,...,M$:
\begin{equation}
(\beta_m, \mathbf{a}_m) = \argmin_{\beta, \mathbf{a}} \sum_{t=1}^{T} (u_{t}-( F_{m-1}(\mathbf{x}_{t}) + \beta h(\mathbf{x}_{t};\mathbf{a}) ))^2
\end{equation}
and
\begin{equation}
F_{m}(\mathbf{x}_{t})= F_{m-1}(\mathbf{x}_{t}) + \beta_m h(\mathbf{x}_{t};\mathbf{a}_m).
\end{equation}
First, the function $h(\mathbf{x}_{t};\mathbf{a})$ is fit by least-squares to the pseudo-residuals $\widetilde{u}_{t,m}$
\begin{equation}
\mathbf{a}_m = \argmin_{\mathbf{a}, \rho} \sum_{t=1}^T [ \widetilde{u}_{t,m} - \rho h(\mathbf{x}_{t};\mathbf{a}) ]^2,
\end{equation}
which at stage $m$ is a residual $\widetilde{u}_{t,m}=(u_{t}-F_{m-1}(\mathbf{x}_t))$. Pseudo-residual at arbitrary stage is defined as
\begin{equation}
\label{eq:pseudo-residual}
\widetilde{u}_{t,m}=-\left[ \frac{\partial \frac{1}{2}(u_{t} - F(\mathbf{x}_t))^2}{\partial F(\mathbf{x}_t)} \right]_{F(\mathbf{x})=F_{m-1}(\mathbf{x})}.
\end{equation}
 The parameter $\rho$ acts as the optimal learning rate in the steepest-descent step, for more details check~\cite{friedman2001greedy}.
Now, we just find the coefficient $\beta_m$ for the expansion as
\begin{equation}
    \beta_m = \argmin_{\beta} \sum_{t=1}^{T} \frac{1}{2}(u_{t}- (F_{m-1}+\beta h(\mathbf{x}_t;\mathbf{a}_m)))^2.
\end{equation}

Each base learner $h(\mathbf{x}_{t};\mathbf{a}_m)$, parameterized with $\mathbf{a}_m$ partitions the feature space $\mathbf{x}_{t} \in \mathbf{X}$
into $L_m$-disjoint regions $\{R_{l,m}\}_1^{L_m}$ and predicts a separate constant value in each:
\begin{equation}
h(\mathbf{x}_{t};\{R_{l,m}\}_1^{L_m})= \sum_{l=1}^{L_m} \bar{u}_{l,m} \mathbf{1}(\mathbf{x}_{t} \in R_{l,m}),
\end{equation}
where $\bar{u}_{l,m}$ is the mean value of pseudo-residual (eq. \ref{eq:pseudo-residual}) in each region $R_{l,m}$
\begin{equation}
\bar{u}_{l,m}=\frac{\sum_{t=1}^T \widetilde{u}_{t,m} \mathbf{1}[\mathbf{x}_t \in R_{l,m}]}{ \sum_{t=1}^T \mathbf{1}[\mathbf{x}_t \in R_{l,m}]}.
\end{equation}
We have used the GBM implementation from Scikit-learn library~\cite{scikit-learn} for all our experiments.
Furthermore, note that different variants of tree boosting have been empirically proven to be state-of-the-art methods in predictive tasks across different machine learning challenges \cite{bentejac2020comparative,chen2016xgboost,lu2020randomized,taieb2014gradient,gulin2011winning} and more recently in finance\cite{zhou2015evolution, sun2018novel}.

\subsection{SGD-based model ensemble}

In stochastic gradient descent (SGD) based optimization, stochasticity comes from two places:
\begin{itemize}
    \item SGD trajectory. 
    The iterates $\{ \Theta(0), \cdots, \Theta(i) \}$ forms a exploratory trajectory, as $\Theta(i)$ is updated by randomly data sample $\mathcal{D}_i$.
    Recent works \cite{mandt2017stochastic, gur2018gradient} studied the connection of trajectory iterates to an approximate Markov chain Monte Carlo sampler by analyzing the dynamics of SGD.
    \item Model initialization. 
    Different initialization of model parameters, i.e. $\Theta(0)$, leads to distinct trajectories.
    It has been shown that ensembles of independently initialized and trained models empirically often provide comparable performance in prediction and uncertainty quantification w.r.t. sampling and variational inference based methods, even though it does not apply conventional Bayesian grounding \cite{lakshminarayanan2017simple, snoek2019can}.
\end{itemize}

In this paper, we make a hybrid approach, that uses both sources of stochasticity to obtain parameter realizations $\{\Theta_m\}$ as follows:
\begin{align}\label{eq:sample}
\{\Theta_m\} \triangleq \bigcup_{j} \{ \Theta^j(i), \cdots, \Theta^j(I) \}
\end{align}
Eq.~\ref{eq:sample} indicates that from each independently trained SGD trajectory (indexed by $j$), we skip the beginning few epochs as a "burn-in" step. 
We choose the remaining as samples from this trajectory. 
Then, we further take the union of samples from independent trajectories as the samples used by the inference in Sec.~\ref{sec:infer}.

In our experiments, we use Adam optimization, a variant of SGD, which has been widely used in machine learning \cite{kingma2014adam}.
We found that $5$ to $25$ independent training processes can give rise to decently accurate and calibrated forecasting.
Moreover, by parallel computing on GPU, we perform each training process in parallel without loss of efficiency.

\subsection{Residual diagnostics of ARMA-GARCH models}

\begin{figure*}[!htbp]
\begin{center}
\piclabC{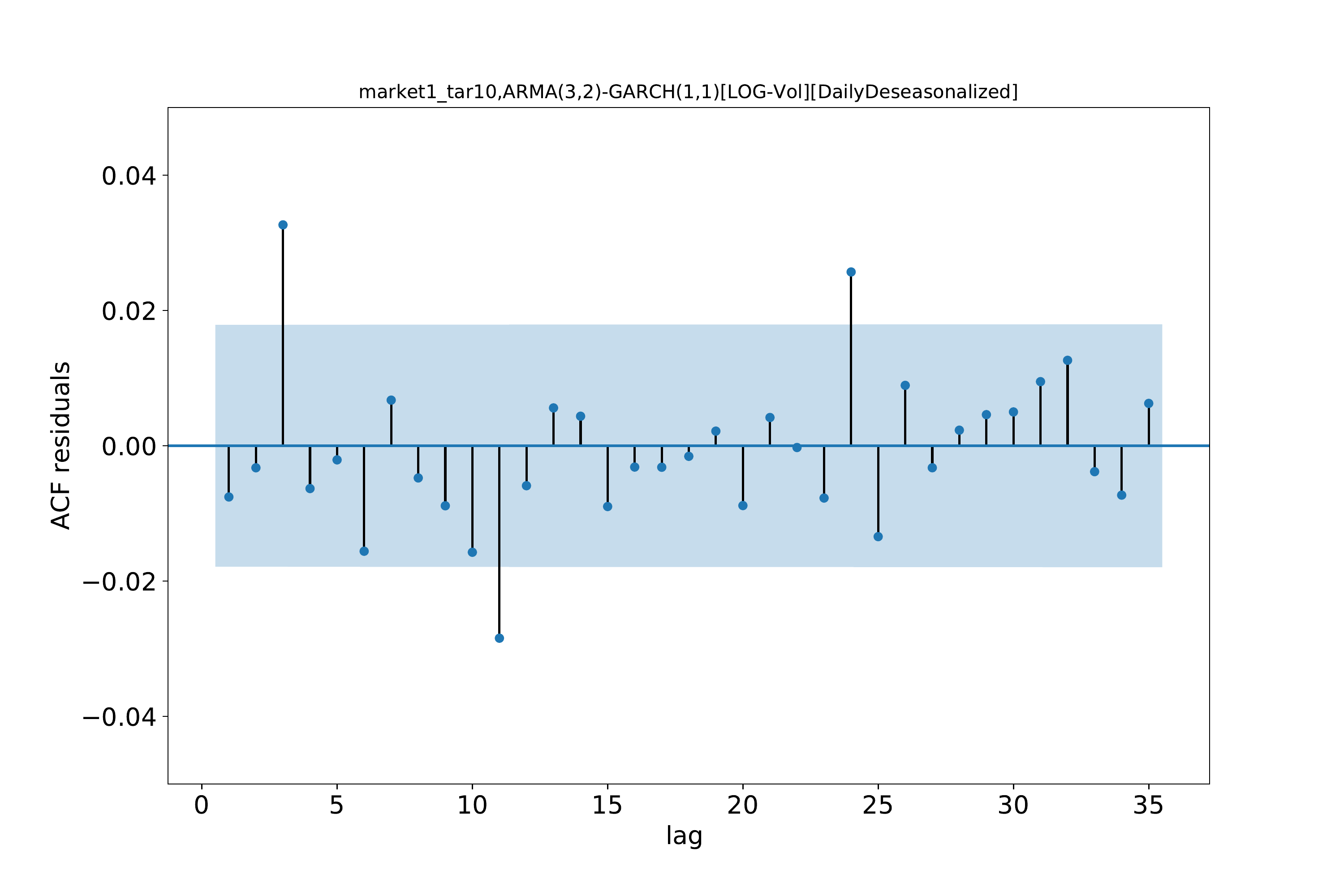}{}
\piclabC{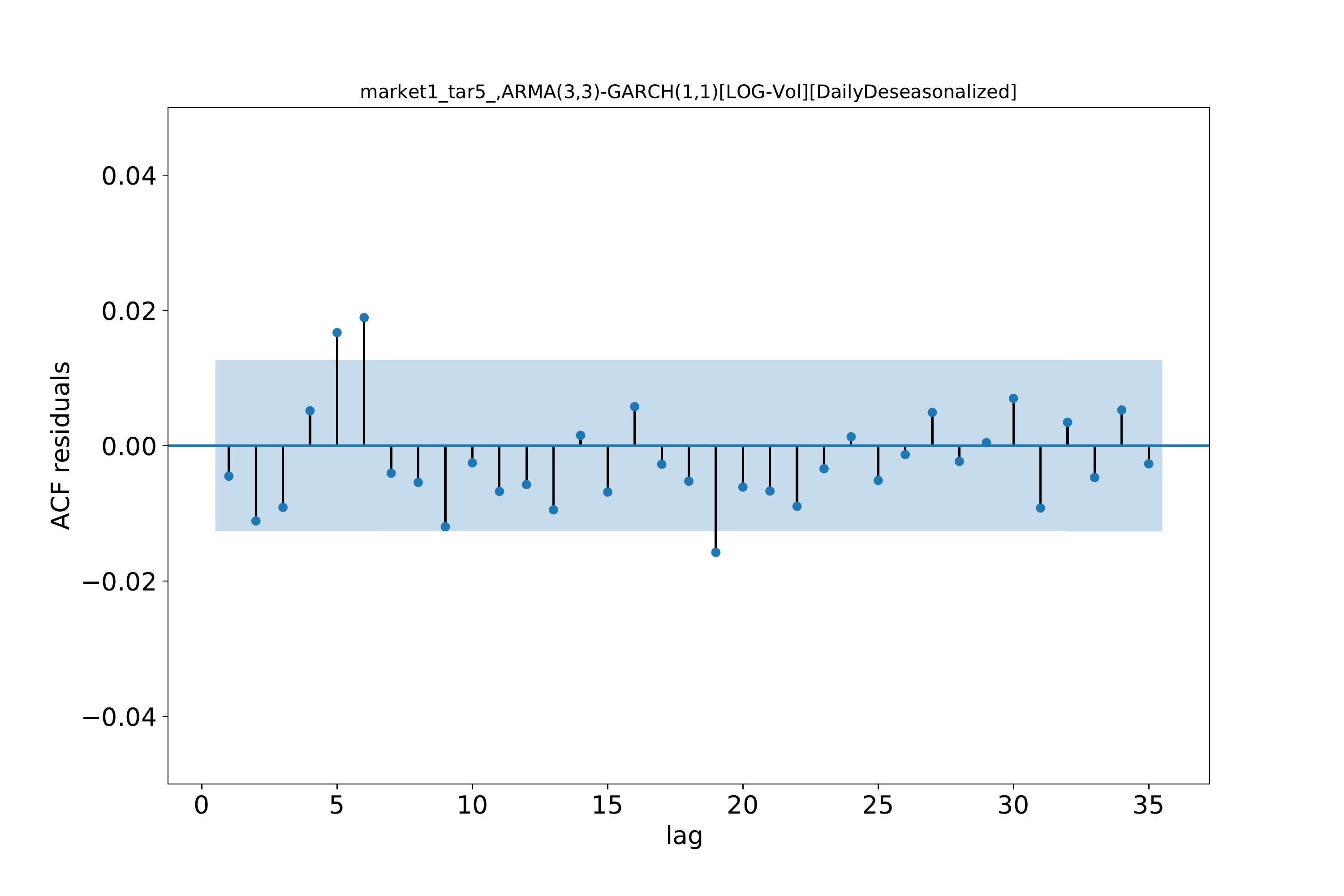}{}
\piclabC{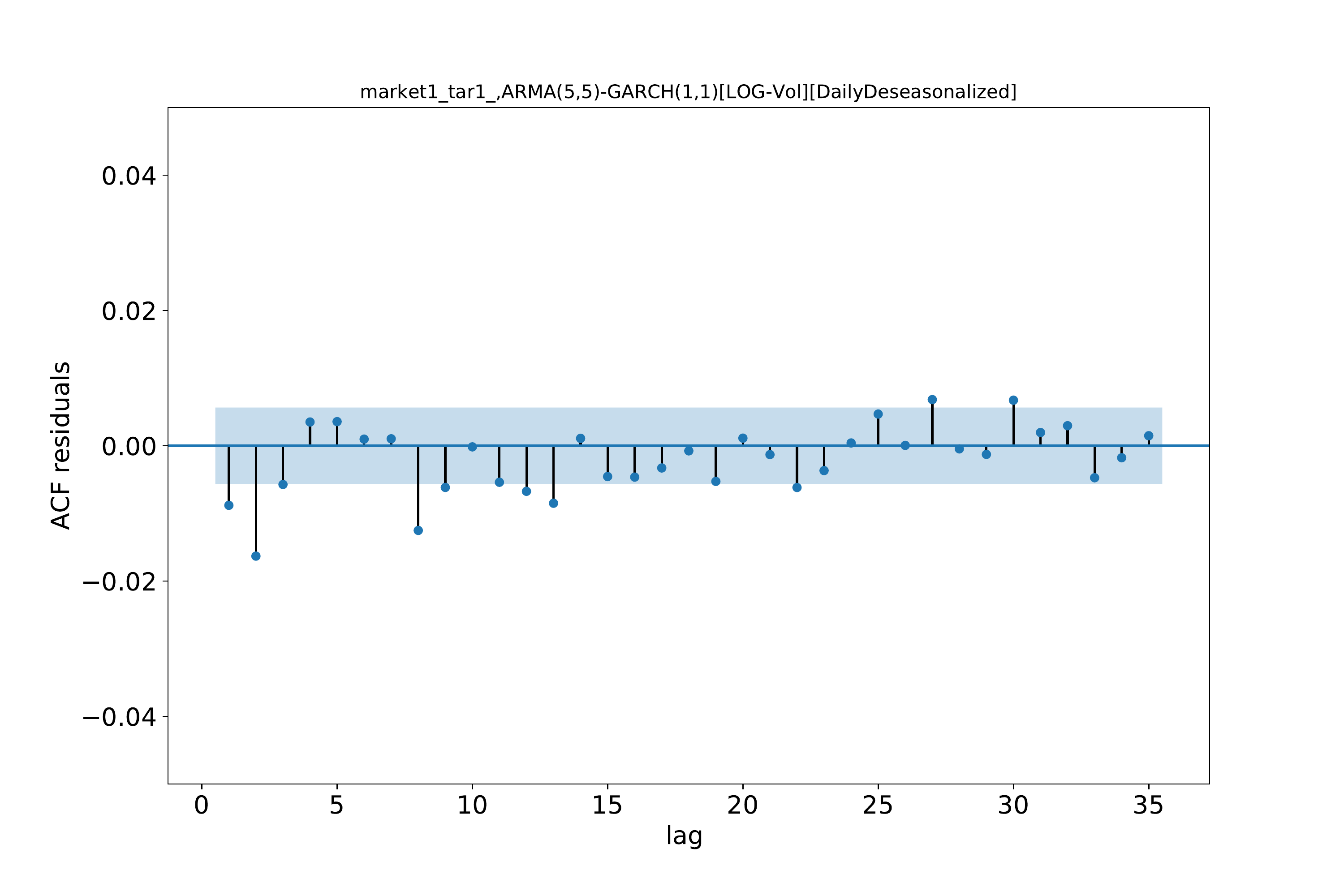}{}
\end{center}
\caption{\textbf{Residual diagnostics} for Bitfinex market: \textbf{First row} ACF of ARMA(3,2)-GARCH(1,1) model residuals on log-volume 10 min.
\textbf{Second row} ACF of ARMA(3,3)-GARCH(1,1) model residuals on log-volume 5 min.
\textbf{Third row} ACF of ARMA(5,5)-GARCH(1,1) model residuals on log-volume 1 min.
}
\label{fig:acf_model_residuals_market1}
\end{figure*}

\begin{figure*}[!htbp]
\begin{center}
\piclabC{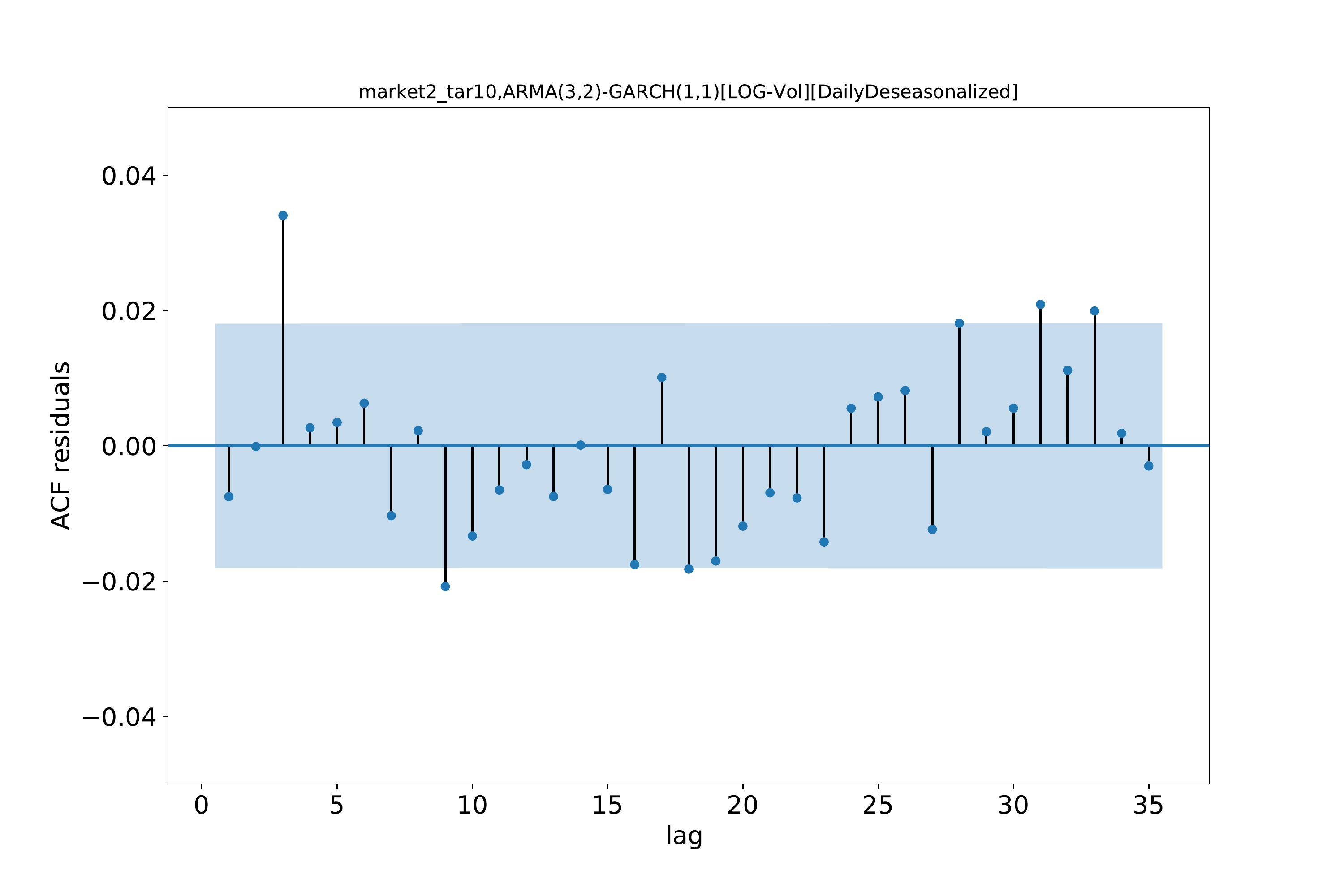}{}
\piclabC{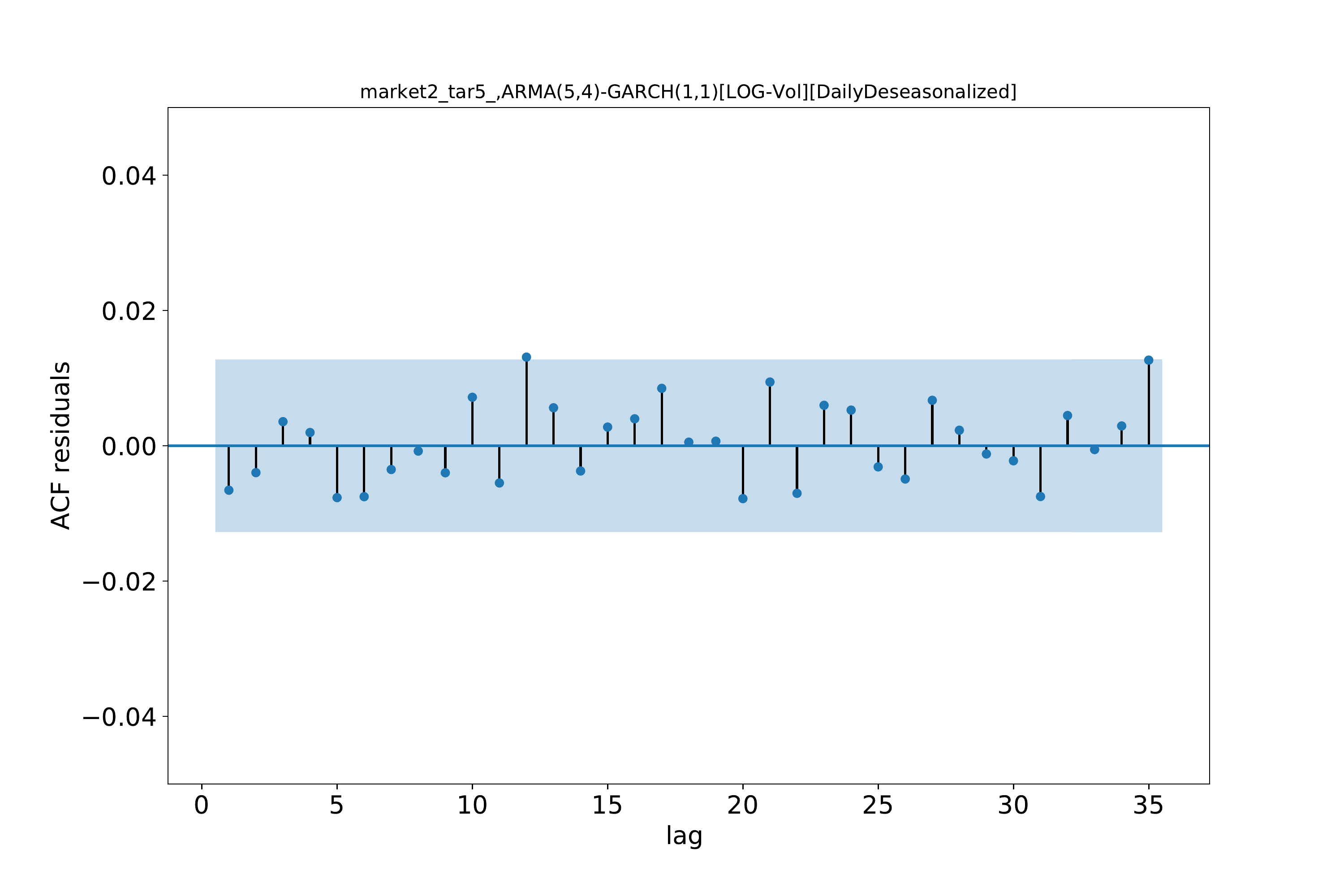}{}
\piclabC{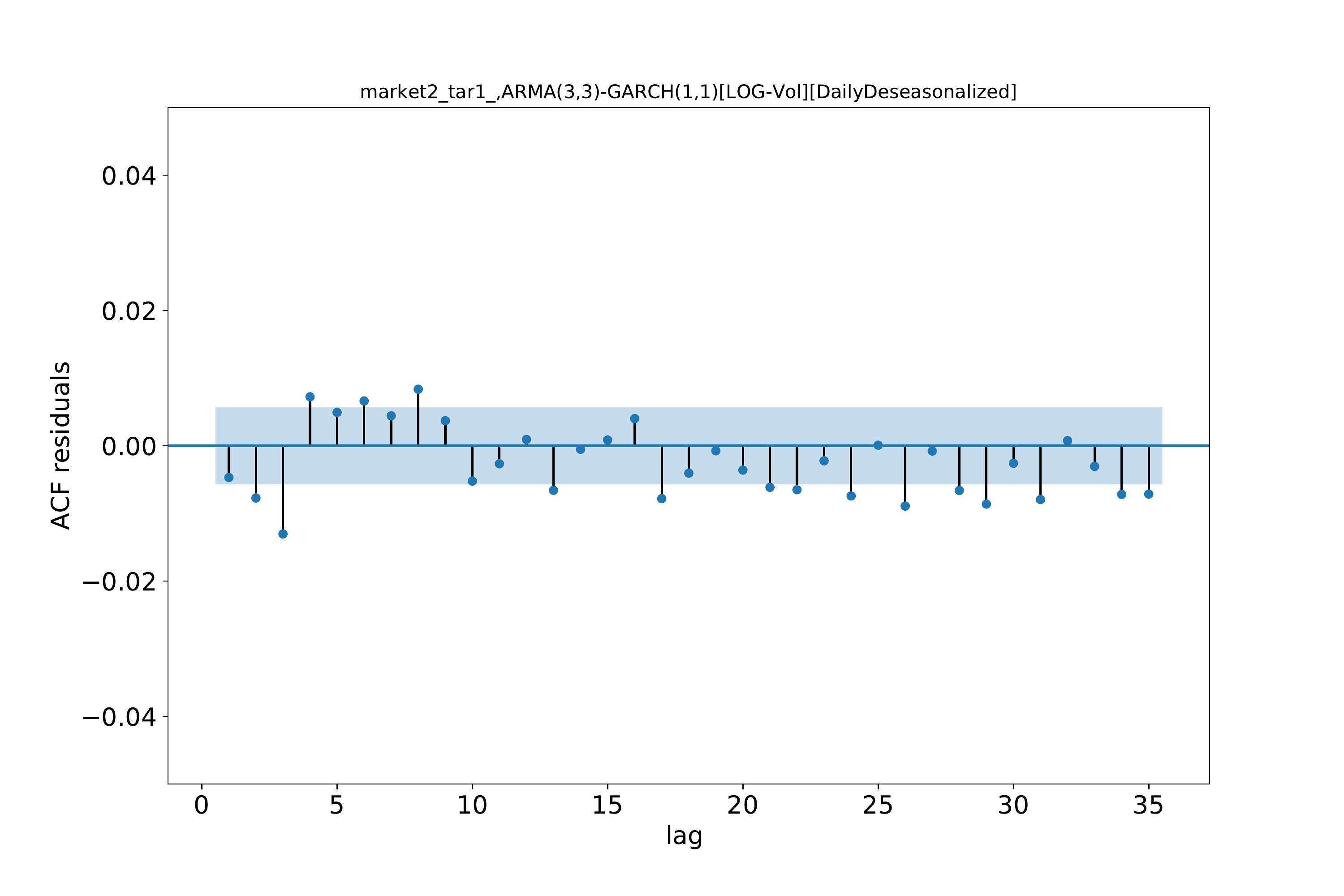}{}
\end{center}
\caption{\textbf{Residual diagnostics} for Bitstamp market: \textbf{First row} ACF of ARMA(3,2)-GARCH(1,1) model residuals on log-volume 10 min.
\textbf{Second row} ACF of ARMA(5,4)-GARCH(1,1) model residuals on log-volume 5 min.
\textbf{Third row} ACF of ARMA(3,3)-GARCH(1,1) model residuals on log-volume 1 min.
}
\label{fig:acf_model_residuals_market2}
\end{figure*}

In Figure \ref{fig:acf_model_residuals_market1} and Figure \ref{fig:acf_model_residuals_market2}, we show auto-correlation function of the residuals for Bitfinex and Bitstamp market, respectively. In all cases of ARMA($p,q$)-GARCH(1,1) models, parameters $p,q$ were fitted to Akaike Information Criterion.  For 1min, 5min and 10min target the majority of residuals ACF are within significance area and only a small number falls very close to the significance area, which implies that the residuals are almost uncorrelated and models well specified.

\newpage

\subsection{Disentangling TME contributions on 1- and 10-min intervals}\label{appA}

\begin{figure}[!htbp]
\centering
\includegraphics[width=0.98\textwidth]{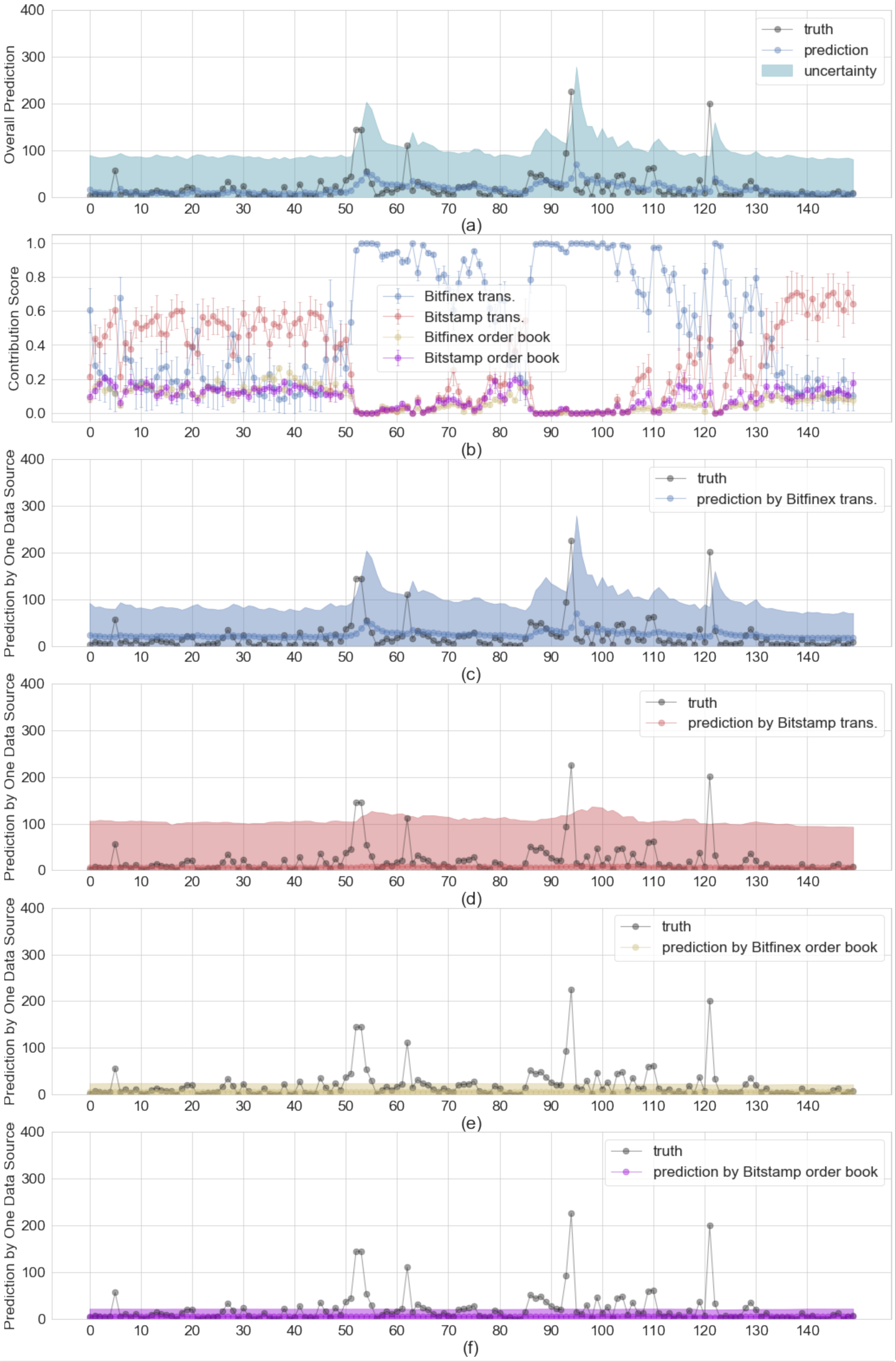}
\caption{Visualization of TME in a sample period of Bitfinex for 1-min volume predicting.
Panel(a): Predictive mean $\pm$ two times standard deviation (left-truncated at zero) and the volume observations.
Panel(b): Data source contribution scores (i.e. average of latent variable probabilities) over time.
Panel(c)-(f): Each data source's predictive mean $\pm$ two times standard deviation (left-truncated at zero). 
The color of each source's plot corresponds to that of the contribution score in Panel(b).
(best viewed in colors)
}
\label{fig:vis_bitfinex1min}
\end{figure}

\begin{figure}[!htbp]
\centering
\includegraphics[width=0.98\textwidth]{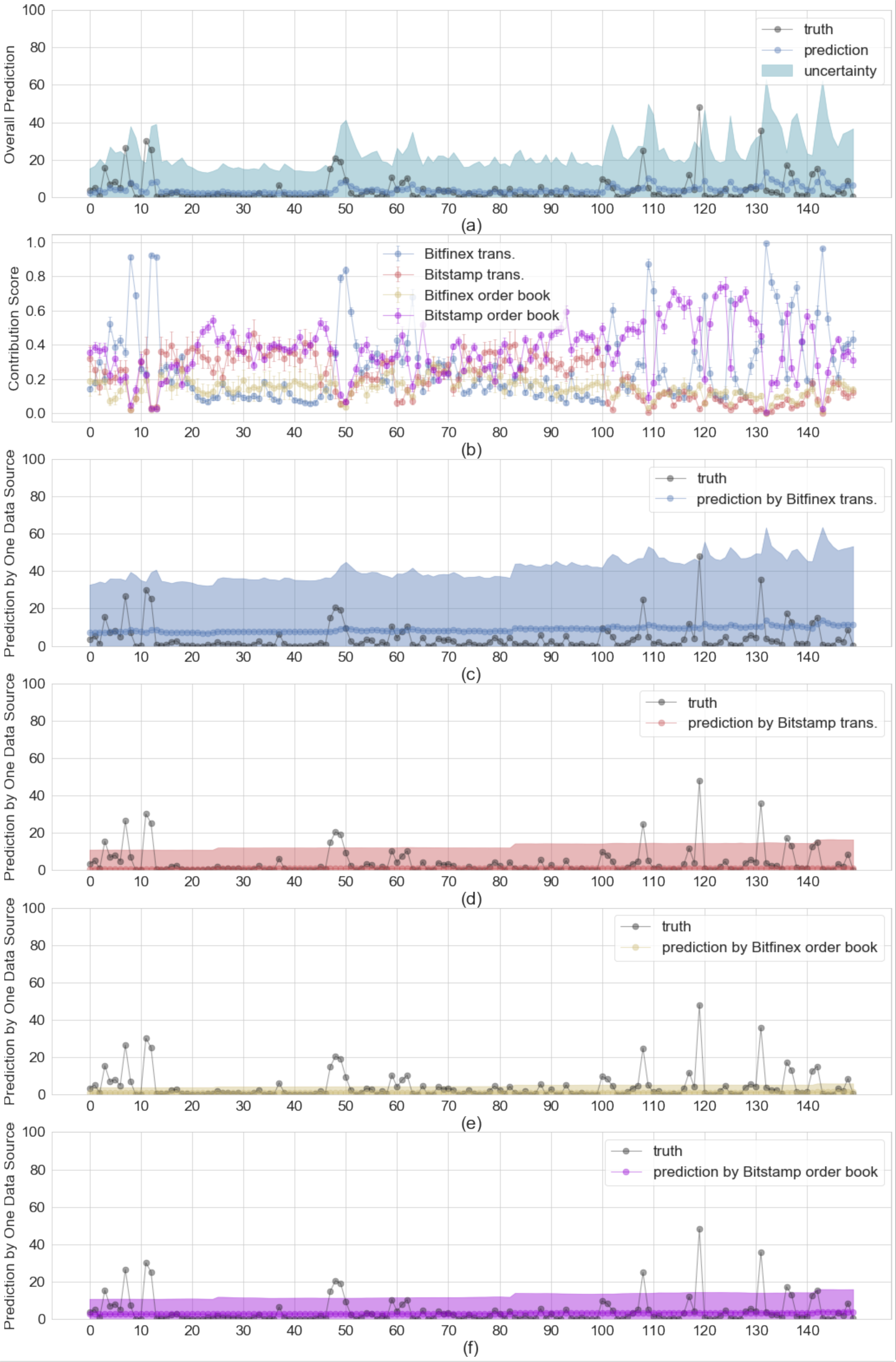}
\caption{Visualization of TME in a sample period of Bitstamp for 1-min volume predicting.
Panel(a): Predictive mean $\pm$ two times standard deviation (left-truncated at zero) and the volume observations.
Panel(b): Data source contribution scores (i.e. average of latent variable probabilities) over time.
Panel(c)-(f): Each data source's predictive mean $\pm$ two times standard deviation (left-truncated at zero). 
The color of each source's plot corresponds to that of the contribution score in Panel(b).
(best viewed in colors)
}
\label{fig:vis_bitstamp1min}
\end{figure}

\begin{figure}[!htbp]
\centering
\includegraphics[width=0.98\textwidth]{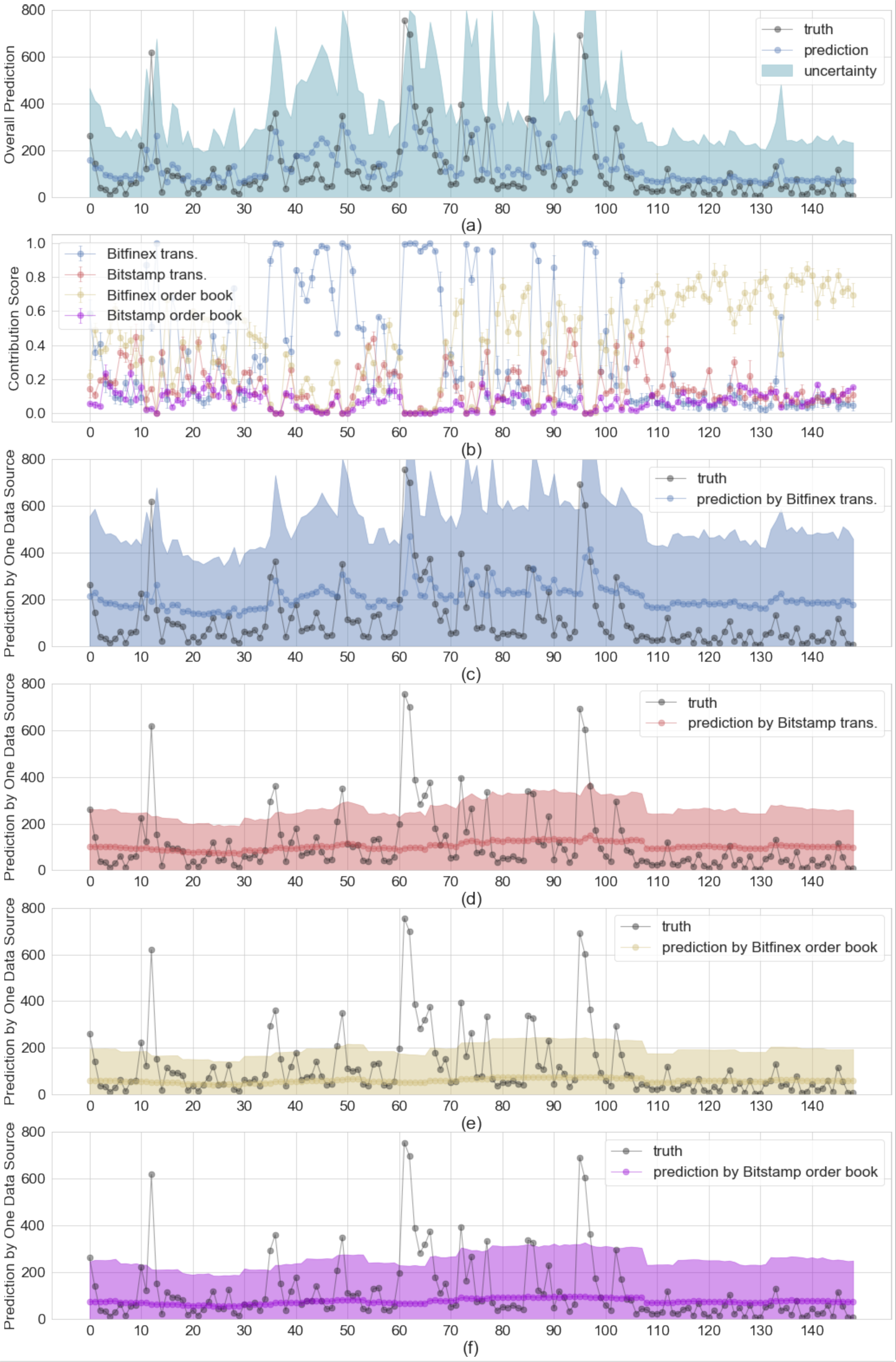}
\caption{Visualization of TME in a sample period of Bitfinex for 10-min volume predicting.
Panel(a): Predictive mean $\pm$ two times standard deviation (left-truncated at zero) and the volume observations.
Panel(b): Data source contribution scores (i.e. average of latent variable probabilities) over time.
Panel(c)-(f): Each data source's predictive mean $\pm$ two times standard deviation (left-truncated at zero). 
The color of each source's plot corresponds to that of the contribution score in Panel(b).
(best viewed in colors)
}
\label{fig:vis_bitfinex10min}
\end{figure}

\begin{figure}[!htbp]
\centering
\includegraphics[width=0.98\textwidth]{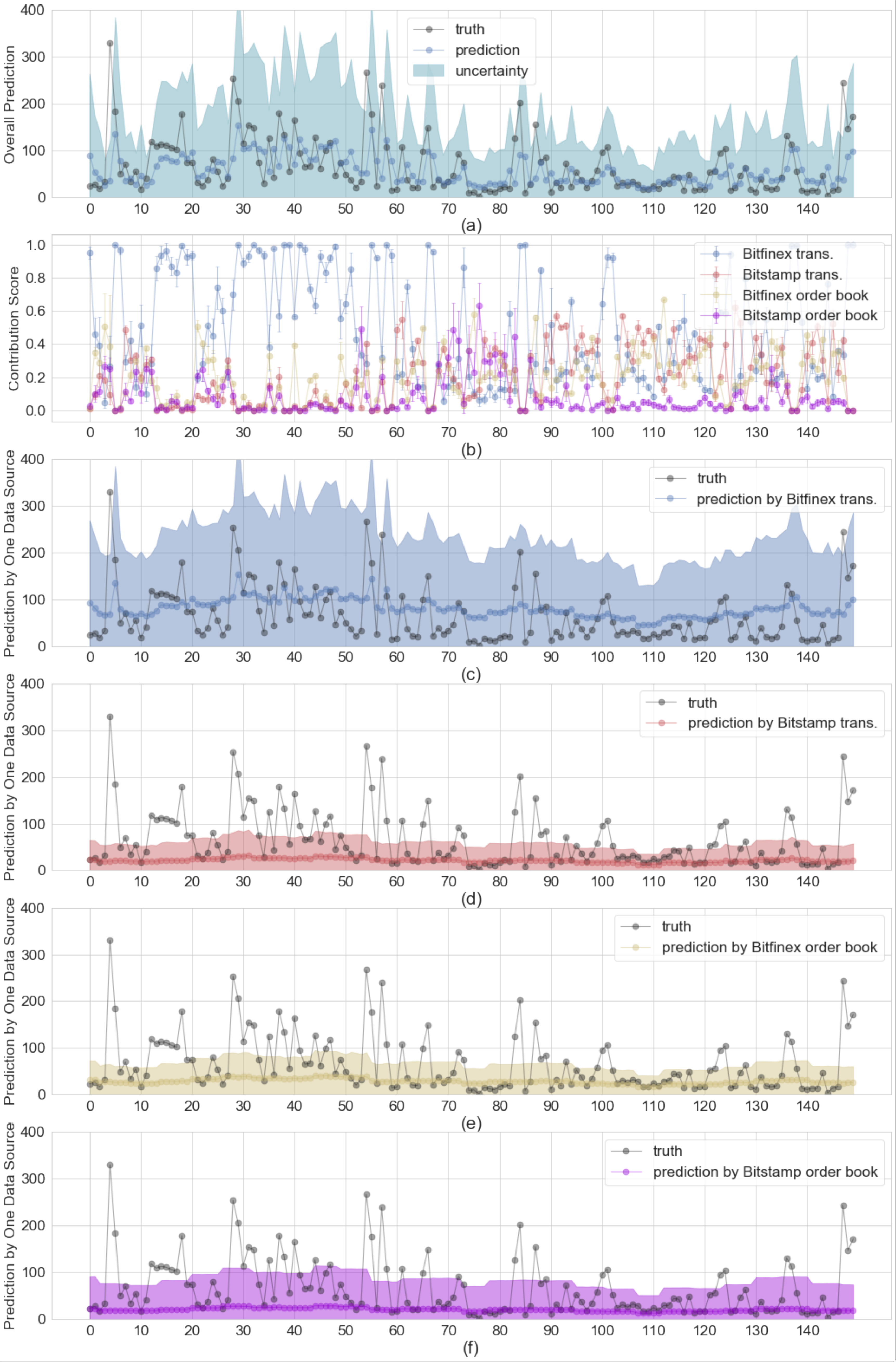}
\caption{Visualization of TME in a sample period of Bitstamp for 10-min volume predicting.
Panel(a): Predictive mean $\pm$ two times standard deviation (left-truncated at zero) and the volume observations.
Panel(b): Data source contribution scores (i.e. average of latent variable probabilities) over time.
Panel(c)-(f): Each data source's predictive mean $\pm$ two times standard deviation (left-truncated at zero). 
The color of each source's plot corresponds to that of the contribution score in Panel(b).
(best viewed in colors)
}
\label{fig:vis_bitstamp10min}
\end{figure}

We report in Figs. \ref{fig:vis_bitfinex1min} and \ref{fig:vis_bitstamp1min} the contributions to TME prediction in the two markets when the sampling interval is 1 minute, while in Figs. \ref{fig:vis_bitfinex10min} and \ref{fig:vis_bitstamp10min} the same figures for 10 min intervals.


\newpage

\subsection{Volume predictions conditional to volume quartile}
 
Table \ref{tab:1minQ1Q4} and \ref{tab:10minQ1Q4} show the performance of the forecast conditioned to volume quartile for the four models and the two markets. As in Table \ref{tab:5minQ1Q4} in the main text, we present both absolute metrics (RMSE and MAE) and relative ones (relative MSE and MAPE).

\begin{table*}[!htbp]
\centering  
  \caption{1-min volume prediction errors conditional on the quantile of the true volume values.
  Measures: RMSE Qx -- root mean squared error conditioned on x-th quantile, RelRMSE Qx -- relative root mean squared error conditioned on x-th quantile, MAE Qx -- mean average error conditioned on x-th quantile, MAPE Qx -- mean absolute percentage error conditioned on x-th quantile. }
  \begin{tabular}{|c|c|c|c|c|}
  \hline
  \textbf{BITFINEX MARKET} & RMSE Q1 & RMSE Q2 & RMSE Q3 & RMSE Q4 \\ 
  \hline 
 ARMA-GARCH  & 13.618         & 17.314         & 20.287         & 38.903 \\
  ARMAX-GARCH & 13.601         & 17.350         & 20.206         & 39.141 \\
  GBM         & \textbf{2.159} & \textbf{2.907} & \textbf{4.323} & 41.671 \\
  TME         &  8.700         & 10.032         & 10.154         & \textbf{36.916} \\
  \hline
  & RelRMSE Q1 & RelRMSE Q2 &  RelRMSE Q3 &  RelRMSE Q4 \\ 
  \hline 
  ARMA-GARCH  & 93998.678          & 16.755         & 4.992          & 1.427\\
  ARMAX-GARCH & 96451.132          & 16.508         & 4.973          & 1.425\\
  GBM         & \textbf{22735.646} & \textbf{2.764} & \textbf{0.968} & 0.754\\
  TME         & 26968.219          & 9.269          & 2.539          & \textbf{0.737}\\
  \hline
  & MAE Q1 & MAE Q2 & MAE Q3 & MAE Q4 \\
  \hline
  ARMA-GARCH  & 9.173          & 11.300         & 12.659         & 23.777 \\
  ARMAX-GARCH & 9.158          & 11.249         & 12.556         & 23.791 \\
  GBM         & \textbf{1.540} & \textbf{1.665} & \textbf{3.054} & 25.652 \\
  TME         & 8.225          &  8.620         &  7.252         & \textbf{20.629} \\
  \hline
  & MAPE Q1 & MAPE Q2 & MAPE Q3 & MAPE Q4 \\
  \hline 
  ARMA-GARCH  & 1319.606         & 10.151         & 2.989          & 0.858\\
  ARMAX-GARCH & 1344.985         & 10.097         & 2.966          & 0.855\\
  GBM         & \textbf{296.446} & \textbf{1.502} & \textbf{0.646} & 0.707\\
  TME         &  318.682         & 7.372          & 1.692          & \textbf{0.568}\\
  \hline
  \hline 
  \textbf{BITSTAMP MARKET} & RMSE Q1 & RMSE Q2 & RMSE Q3 & RMSE Q4 \\ 
  \hline
  ARMA-GARCH    & 8.2704 & 8.9885 & 11.0034 & 24.1002\\
  ARMAX-GARCH  & 7.9412 & 8.6546 & 10.5348 & 23.8335\\
  GBM          & \textbf{0.929} & \textbf{1.085}  & \textbf{1.721} &   23.373 \\
  TME          &    3.047       &   3.426         &    3.540       & \textbf{22.005} \\
  \hline
   & RelRMSE Q1 & RelRMSE Q2 &  RelRMSE Q3 &  RelRMSE Q4 \\ 
  \hline 
  ARMA-GARCH  & 16842.969         & 27.709         & 6.423          & 1.974\\
  ARMAX-GARCH & 16501.630         & 26.679         & 6.135          & 1.923\\
  GBM         & \textbf{2082.545} & \textbf{3.190} & \textbf{0.845} & 0.786\\
  TME         &  7279.497         & 10.951         & 2.107          & \textbf{0.696} \\
  \hline
  & MAE Q1 & MAE Q2 & MAE Q3 & MAE Q4 \\
  \hline
  ARMA-GARCH    & 5.400           & 6.125           & 7.003            & 12.223\\
  ARMAX-GARCH   & 5.239           & 5.944           & 6.778            & 11.985\\
  GBM           & \textbf{0.599}  & \textbf{0.605}  & \textbf{1.316}   & 11.540\\
  TME           & 2.664           & 2.780           & 2.259            & \textbf{9.615}\\
  \hline
  & MAPE Q1 & MAPE Q2 & MAPE Q3 & MAPE Q4 \\
  \hline 
  ARMA-GARCH   & 1043.922         & 17.211         & 3.891          & 1.097\\
  ARMAX-GARCH  & 1018.872         & 16.694         & 3.767          & 1.066\\
  GBM          & \textbf{118.256} & \textbf{1.720} & \textbf{0.632} & 0.741\\
  TME          &  500.100         &  8.035         & 1.301          & \textbf{0.571}\\
  \hline
\end{tabular}
\label{tab:1minQ1Q4}
\end{table*}

\begin{table*}[htbp]
\centering
  \caption{10-min volume prediction errors conditional on the quartile of the true volume values.
  Measures: RMSE Qx -- root mean squared error conditioned on x-th quantile, RelRMSE Qx -- relative root mean squared error conditioned on x-th quantile, MAE Qx -- mean average error conditioned on x-th quantile, MAPE Qx -- mean absolute percentage error conditioned on x-th quantile.}
  \begin{tabular}{|c|c|c|c|c|}
  \hline
  \textbf{BITFINEX MARKET} & RMSE Q1 & RMSE Q2 & RMSE Q3 & RMSE Q4 \\
  \hline 
  ARMA-GARCH  & 61.494 & 70.540 & 74.188 & 191.332\\
  ARMAX-GARCH &  59.438 & 67.474 & 71.237 & 192.155\\
  GBM         & \textbf{39.380} & \textbf{39.576} & \textbf{46.774} & 207.957 \\
  TME         & 57.252          & 64.705          & 67.354          & \textbf{185.136} \\
  \hline
  & RelRMSE Q1 & RelRMSE Q2 & RelRMSE Q3 & RelRMSE Q4 \\
  \hline 
  ARMA-GARCH  & 9.262          & 1.998          & 1.037          & 0.544 \\
  ARMAX-GARCH & 8.822          & 1.911          & 0.990          & \textbf{0.538} \\
  GBM         & \textbf{6.154} & \textbf{1.136} & \textbf{0.587} & 0.571 \\
  TME         &11.434          & 1.878          & 0.945          & 0.543\\
  \hline
  & MAE Q1 & MAE Q2 & MAE Q3 & MAE Q4 \\
  \hline
  ARMA-GARCH & 50.542 & 54.148 & 53.169 & 131.891\\
  ARMAX-GARCH & 48.985 & 52.116 & 51.257 & 132.081\\
  GBM        & \textbf{33.788} & \textbf{28.760} & \textbf{34.892} & 148.550  \\
  TME        & 46.680          & 48.797          & 43.890          & \textbf{128.578} \\
  \hline
   & MAPE Q1 & MAPE Q2 & MAPE Q3 & MAPE Q4 \\
  \hline
  ARMA-GARCH   & 5.109          & 1.503          & 0.705          & 0.462\\
  ARMAX-GARCH  & 4.938          & 1.448          & 0.678          & \textbf{0.461}\\
  GBM          & \textbf{3.561} & \textbf{0.804} & \textbf{0.434} & 0.518\\
  TME          & 6.739          & 1.499          & 0.590          & 0.465\\
  \hline
  \hline
  \textbf{BITSTAMP MARKET} & RMSE Q1 & RMSE Q2 & RMSE Q3 & RMSE Q4 \\
  \hline
  ARMA-GARCH  & 25.000 & 27.672 & 27.541 & 124.527\\
  ARMAX-GARCH & 24.963 & 29.748 & 29.545 & 126.985\\
  GBM         & \textbf{14.355} & \textbf{14.700} & \textbf{16.749} & 131.497\\
  TME         & 25.469          & 27.221          & 24.905          & \textbf{123.804}\\
  \hline
  & RelRMSE Q1 & RelRMSE Q2 & RelRMSE Q3 & RelRMSE Q4 \\
  \hline
  ARMA-GARCH  & 7.210          & 1.860          & 0.949          & 0.619\\
  ARMAX-GARCH & 6.926          & 1.974          & 0.984          & 0.739\\
  GBM         & \textbf{5.638} & \textbf{0.993} & \textbf{0.541} & 0.625\\
  TME         &12.308          & 1.838          & 0.858          & \textbf{0.574}\\
  \hline
  & MAE Q1 & MAE Q2 & MAE Q3 & MAE Q4 \\
  \hline
  ARMA-GARCH  & 20.014 & 21.343 & 19.398 & 66.899\\
  ARMAX-GARCH & 19.633 & 21.685 & 19.718 & 70.021\\
  GBM         & \textbf{11.622} & \textbf{10.490} & \textbf{12.965} & 75.674 \\
  TME         & 21.233          & 20.034          & 17.787          & \textbf{64.918} \\
  \hline
  & MAPE Q1 & MAPE Q2 & MAPE Q3 & MAPE Q4 \\
  \hline
  ARMA-GARCH   & 4.866          & 1.410          & 0.646          & 0.487\\
  ARMAX-GARCH  & 4.716          & 1.429          & 0.653          & 0.528\\
  GBM          & \textbf{3.341} & \textbf{0.695} & \textbf{0.414} & 0.555\\
  TME          & 7.205          & 1.457          & 0.550          & \textbf{0.478}\\
  \hline
\end{tabular}
\label{tab:10minQ1Q4}
\end{table*}

\end{document}